\documentclass[12pt]{article}
\usepackage{amsmath,amssymb,amsthm,pstricks,pst-plot,pst-node}
\usepackage[dvips]{graphics}

\newpsobject{showgrid}{psgrid}{subgriddiv=1,griddots=5,gridlabels=0pt}

\newtheorem{thm}{Theorem}[section]
\newtheorem{lem}[thm]{Lemma}
\newtheorem{cor}[thm]{Corollary}
\newtheorem{prop}[thm]{Proposition}

\newcommand{\R}{{\mathbb R}}

\newcommand{\E}{{\mathbb E}}
\newcommand{\Eng}{{\cal E}}
\newcommand{\C}{{\mathbb C}}
\newcommand{\Z}{{\mathbb Z}}
\newcommand{\M}{{\cal M}}
\newcommand{\T}{{\mathbb T}}

\newcommand{\A}{{\mathbb A}}

\newcommand{\Cov}{{\rm Cov}}

\renewcommand{\Re}{{\rm{Re}}}
\renewcommand{\Im}{{\rm{Im}}}

\input epsf.tex

\newcommand{\bb}{\mathsf{b}}
\newcommand{\ww}{\mathsf{w}}

\begin{document}
\title{Dimers and Amoebae}
\author{Richard Kenyon
\thanks{Laboratoire de Math\'ematiques UMR 8628 du CNRS, B\^at 425, Universit\'e Paris-Sud, 91405
Orsay, France.} \and Andrei Okounkov \thanks{Princeton University, Department of
Mathematics, Princeton, NJ 08544, U.S.A.}
 \and Scott Sheffield
\thanks{Stanford University and Microsoft Research}} \date{} \maketitle

\begin{abstract} We study random surfaces which arise as height functions of random perfect
matchings (a.k.a.\  dimer configurations) on an weighted, bipartite, doubly periodic graph $G$
embedded in the plane. We derive explicit formulas  for the surface tension
 and local Gibbs measure probabilities of these models. The answers involve
a certain plane algebraic curve, which is
the spectral curve of the Kasteleyn operator of the graph. For example,
the surface tension is the
Legendre dual of the Ronkin function of the spectral curve. The
amoeba of the spectral curve represents the phase diagram of the
dimer model. Further, we prove that the spectral curve
of a dimer model is always a
real curve of special type, namely it is a Harnack curve. This
implies many qualitative and quantitative statement about the behavior of
the dimer model, such as existence of smooth phases, decay rate
of correlations, growth rate of
height function fluctuations, etc.
\end{abstract}

\tableofcontents

\section{Introduction}

A \emph{perfect matching} of a graph is a collection of edges with the property that each vertex is
incident to
exactly one of these edges. A graph is \emph{bipartite} if the vertices can be
$2$-colored, that is, colored black and white so that black vertices are adjacent only to white
vertices and vice versa.

Random perfect matchings of a planar graph $G$---also called \emph{dimer configurations}---
are sampled uniformly (or alternatively, with a probability proportional to a product of 
the corresponding edge
weights of $G$) from the set of all perfect matchings on $G$.  These so-called dimer models are the
subject of an extensive physics and mathematics literature.  
(See \cite{Kenyon.surv} for a survey.)

\begin{figure}[htbp]
\begin{center}\scalebox{0.52}{\leavevmode \epsfbox[20 5 300 440]{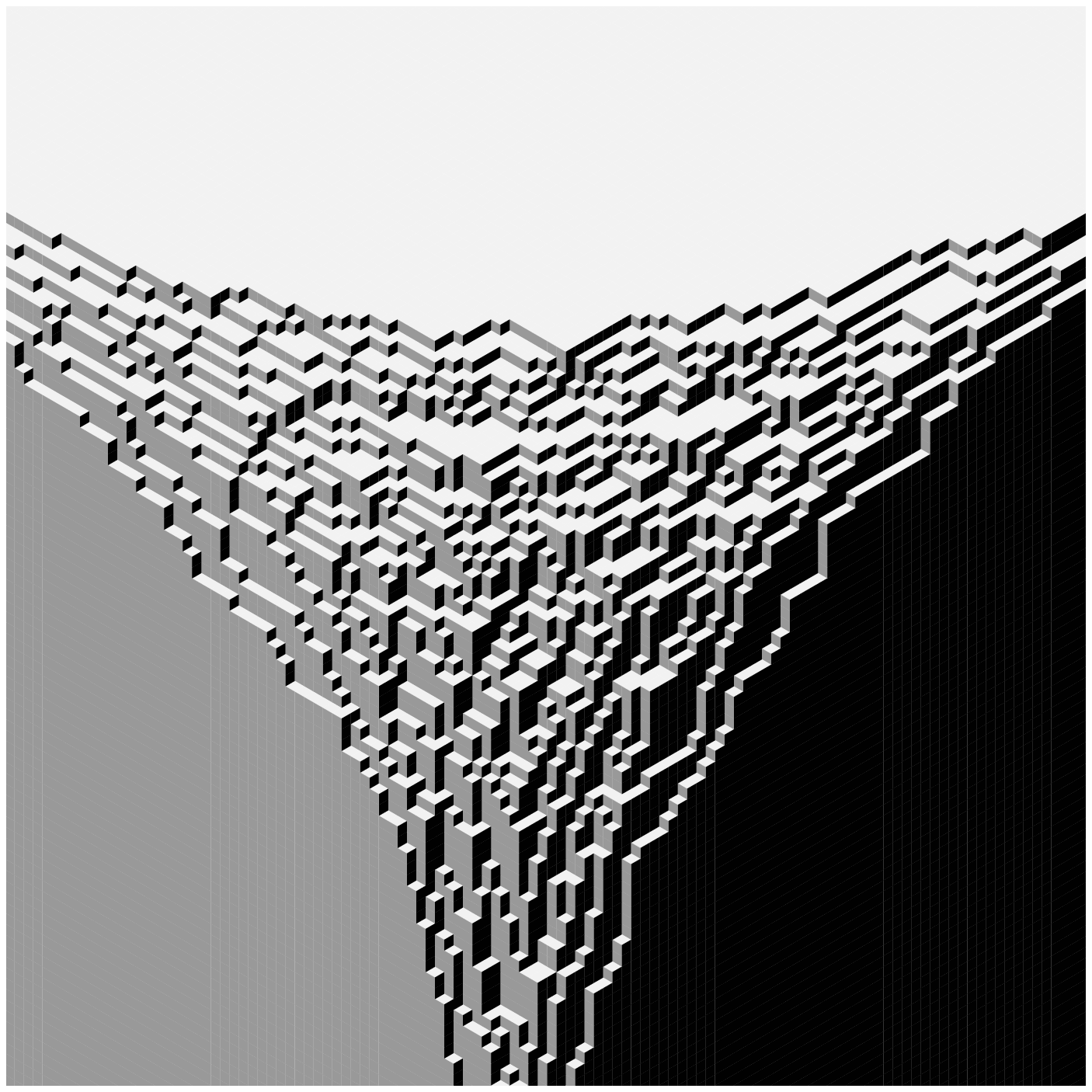}}  
\hfill 
\scalebox{0.55}{\includegraphics{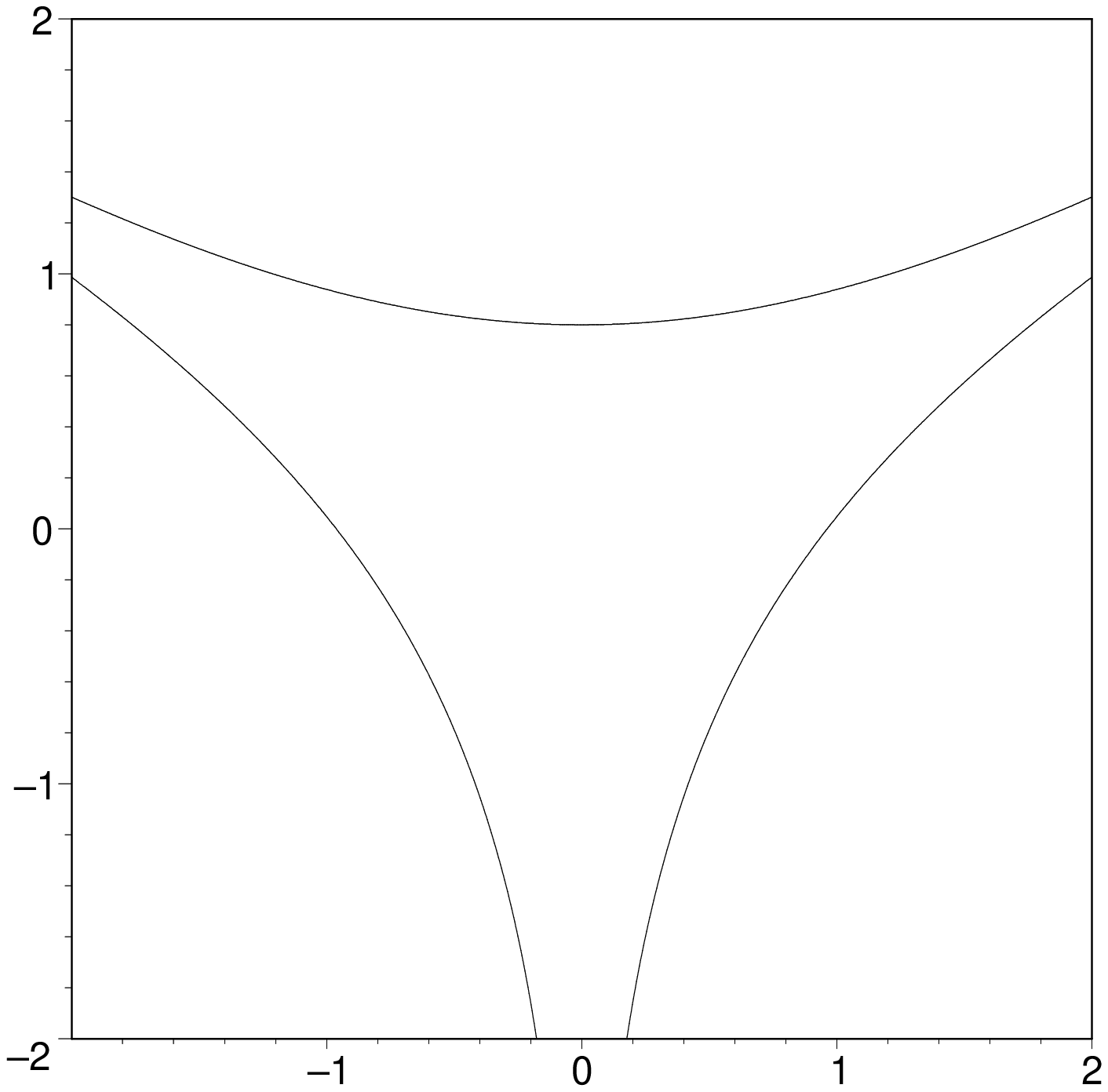}}
\end{center}
\vspace{-3 mm} 
\caption{\label{3DY}\sl On the left is the height function of a 
random volume-constrained dimer configuration on 
the honeycomb lattice. The boundary conditions here are that of
a crystal corner: all dimers are aligned the same way deep enough in each of the 3 sectors.
On the right is (the boundary of) the amoeba of a straight line.}
\end{figure}

Since the set of perfect matchings of $G$ is also in one-to-one correspondence with a class of
height functions on the faces of $G$, we may think of random perfect matchings as (discretized)
\emph{random surfaces}.  One reason for the
interest in perfect matchings is that random surfaces of
this type (and a more general class of random surfaces called \emph{solid-on-solid models}) are
popular models for crystal surfaces (e.g.\ partially dissolved salt crystals) at equilibrium. These
height functions are most visually compelling when $G$ is a honeycomb lattice.  
In this case, we may represent the vertices of $G$ by triangles in a triangular 
lattice and edges of $G$ by rhombi formed by two adjacent triangles. Dimer 
configurations correspond to tilings by such rhombi; they can be viewed as 
planar projections of surfaces of the kind seen in Figure \ref{3DY}.
The third coordinate, which can be reconstructed from the dimer
configuration uniquely, up to an overall additive constant, is the height function. 

Most random surface models  cannot be solved exactly, and
authors are content to prove qualitative results about the surface tension, the existence of
facets, the set of gradient Gibbs measure, etc.

We will prove in this paper, however, that models based on perfect matchings (on any weighted
doubly-periodic bipartite graph $G$ in the plane) are exactly solvable in a rather strong sense.
Not only can we derive explicit formulas for the surface tension---we also explicitly classify the set of Gibbs measures on tilings and
explicitly compute the local probabilities in each of them.  These results are a generalization of
\cite{CKP} where similar results for
$G= \mathbb Z^2$ with constant edge weights were obtained.

The theory has some surprising connections to algebraic geometry.  In
particular, the phase diagram of the Gibbs measures for dimers on a particular graph is
represented by the  \emph{amoeba} of an associated plane algebraic curve,
the \emph{spectral curve}, see Theorem \ref{mainmsr}. 
We recall that by definition \cite{GKZ,Mikhalkin} the
amoeba of an affine algebraic variety $X\in \C^n$
 (plane curve, in our case) is the image of $X$ under the
map taking coordinates to the logarithms of their absolute value. See Figure
\ref{am4by4} for an illustration. The so-called \emph{Ronkin function}
of the spectral curve turns out to be the Legendre dual of the surface tension
(Theorem \ref{legendrethm}).

\begin{figure}[htbp]
\begin{center}\scalebox{0.52}{\leavevmode \epsfbox[20 5 300 440]{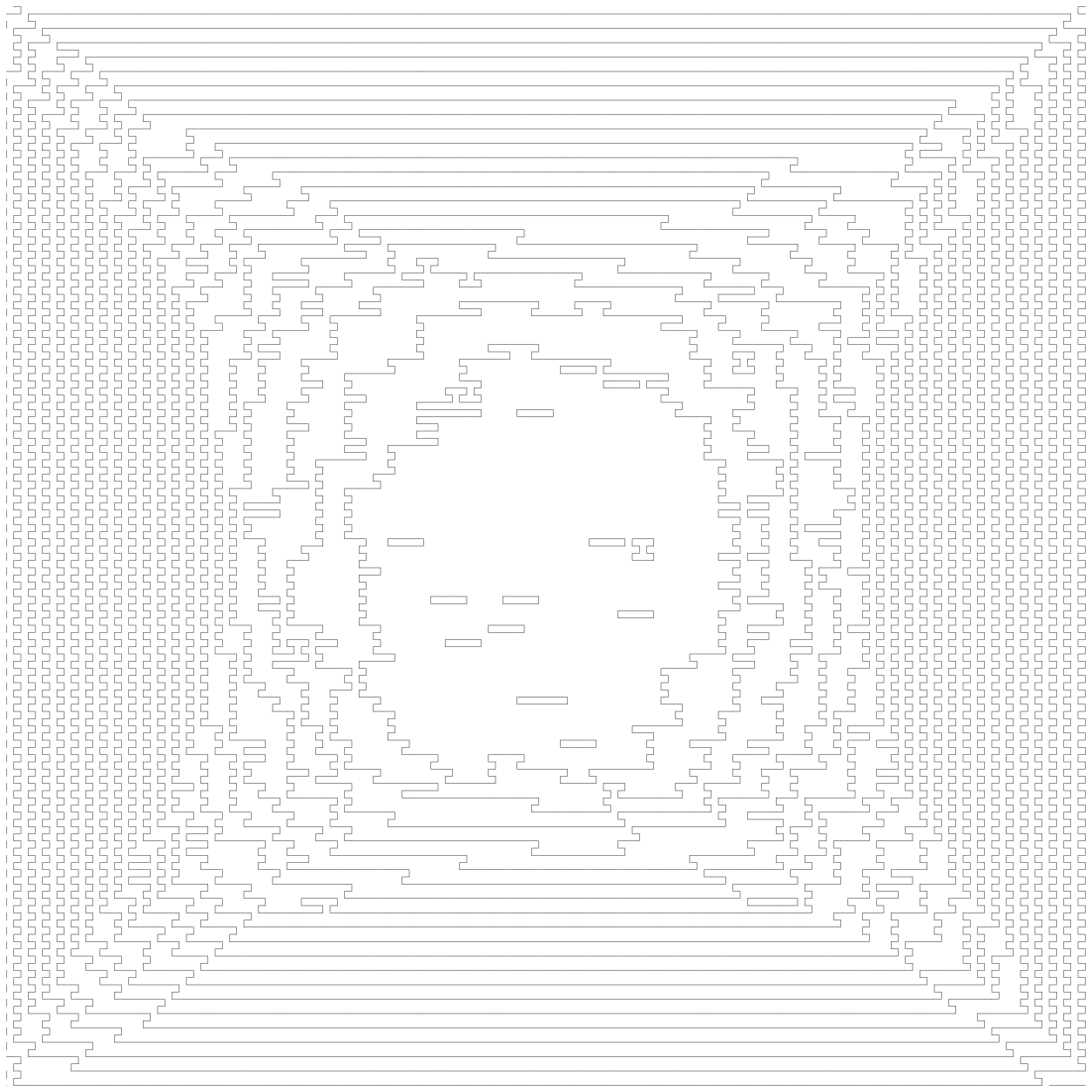}}  
\hfill 
\scalebox{0.55}{\includegraphics{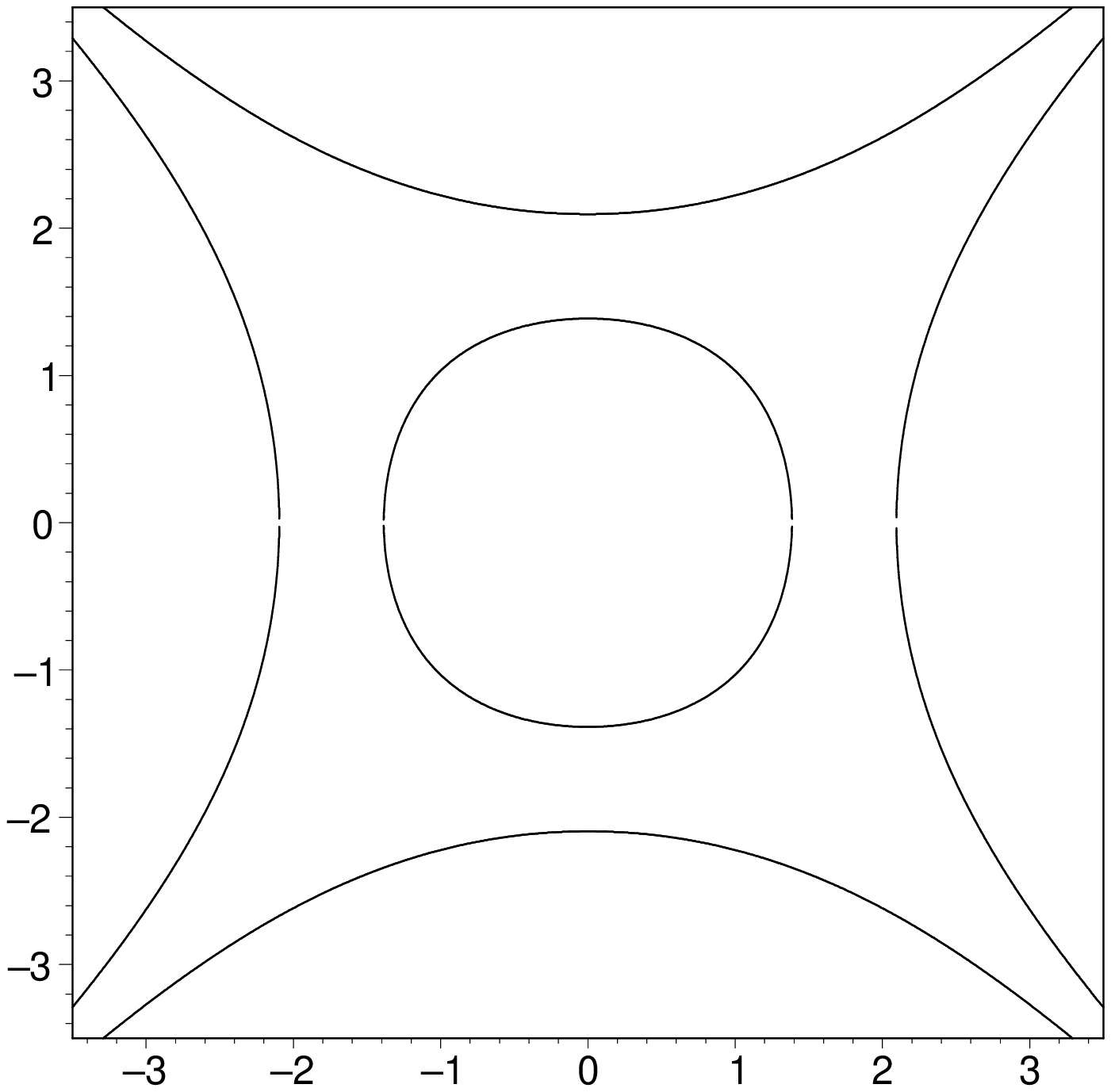}}
 \end{center}
\vspace{-3 mm} 
\caption{\label{fsimul}\sl On the left are the 
level sets of perimeter 10 or longer of 
the height function of a random volume-constrained 
dimer configuration on $\Z^2$ (with $2\times 2$ 
fundamental domain). 
The height function is essentially constant in the middle --- a facet
is developing there. The intermediate region, in which the 
height function is not approximately linear, converges to the 
amoeba of the spectral curve, which can be seen on the 
right. The spectral curve in this case is a genus $1$ curve
with the equation $z+z^{-1}+w+w^{-1}=6.25$.}
\end{figure}

Crystal facets in the
model are in bijection with the components of the
complement of the amoeba. In particular, the 
bounded ones correspond to compact holes in 
the amoeba; the number of bounded facets equals the 
genus of the spectral curve. By the Wulff construction, 
the Ronkin function describes the limit height
function in the dimer model for suitable boundary
conditions (which can be interpreted as the shape of 
a partially dissolved crystal corner). For example, the limit shape 
in the situation shown in Figure \ref{3DY} is the Ronkin function of the 
straight line. It has genus zero and, hence, has no 
bounded facets. A more complicated limit shape,
in which a bounded facet develops, can be seen in Figures 
\ref{fsimul} and \ref{fRonk}. 
\vspace{-5 mm} 

Crystals that appear in nature typically have a small number of facets---the slopes of which are
rational with respect to the underlying crystal lattice.  But laboratory conditions have produced
equilibrium surfaces with up to sixty identifiably different facet slopes \cite{P}.  It is
therefore of interest to have a model in which it is possible to generate crystal surfaces with
arbitrarily many facets and to observe precisely how the facets evolve when weights and temperature
are changed.

For another surprising connection between dimers and algebraic geometry
see \cite{ORV}.

\bigskip
\noindent
\textbf{Acknowledgments}

\nobreak

\medskip
\noindent
The paper was completed while R.~K.\ was visiting Princeton University. 
A.~O.\ was partially supported by DMS-0096246 and a fellowship from
Packard foundation.

\section{Definitions}
\subsection{Combinatorics of Dimers}
\subsubsection{Periodic bipartite graphs and matchings} 
\begin{figure}[!hbtp]
  \begin{center}
    \scalebox{1}{\includegraphics{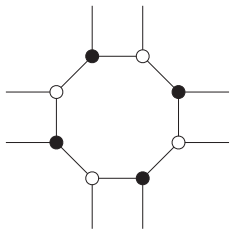}}
    \caption{\sl The fundamental domain of the square-octagon graph.}
    \label{fsqoct}
  \end{center}
\end{figure}

Let $G$ be a $\Z^2$-periodic bipartite
planar graph. By this we mean $G$ is embedded in the plane so that translations in $\Z^2$ act by
color-preserving isomorphisms of $G$ -- isomorphisms which map black vertices to black vertices and
white to white. An example of such graph is the square-octagon 
graph, the fundamental domain of which is shown in Figure
\ref{fsqoct}. 
More familiar (and, in a certain precise sense, universal)
examples are the standard square and honeycomb lattices. 
Let $G_n$ be the quotient of $G$ by the action of $n\Z^2$. It is a finite bipartite
graph on a torus. 

Let $\M(G)$ denote the set of perfect matchings of $G$.  A well-known necessary and sufficient
condition for the existence of a perfect matching of $G$ is the existence of a unit flow from white
vertices to black vertices, that is a flow with source $1$ at each white vertex and sink $1$ at
every black vertex.  If a unit flow on $G$ exists, then by taking averages of the flow over larger
and larger balls and subsequential limits one obtains a unit flow on $G_1$. Conversely, if a unit
flow on $G_1$ exists, it can be extended periodically to $G$.  Hence $G_1$ has a perfect matching
if and only if $G$ has a perfect matching.

\subsubsection{Height function}
Any matching $M$ of $G$ defines a white-to-black unit flow $\omega$:
flow by one along each matched edge.
Let $M_0$ be a fixed periodic matching of $G$ and $\omega_0$ the corresponding flow.
For any other matching $M$ with flow $\omega$,
the difference $\omega-\omega_0$ is a divergence-free flow. Given
two faces $f_0,f_1$ let $\gamma$ be a path in the dual graph $G^*$ from $f_0$
to $f_1$. The total flux of $\omega-\omega_0$ across $\gamma$
is independent of $\gamma$ and therefore is a function of $f_1$ called the \emph{height function} of $M$.

The height function of a matching $M$ is well-defined
up to the choice of a base face $f_0$
and the choice of reference matching $M_0$.
The difference of the height functions of two matchings is well-defined
independently of $M_0$.

A matching $M_1$ of $G_1$ defines a periodic matching $M$ of $G$;
we say $M_1$ has \emph{height change $(j,k)$} if the
horizontal and vertical height
changes of $M$ for one period are $j$ and $k$ respectively, that is
$$
h(v+(x,y))=h(v)+jx+ky
$$ where $h$ is the height function on $M$. The height change
is an element of $\Z^2$, and can be identified with the homology class
in $H_1(\T^2,\Z)$ of the flow $\omega_1-\omega_0$.

\subsection{Gibbs measures}
\subsubsection{Definitions}
Let $\Eng$ be a real-valued function on edges of $G_1$, the
\emph{energy} of an edge. It defines a periodic energy function
on edges of $G$. We define the energy of a finite set $M$ of edges
by $\Eng(M)=\sum_{e\in M}\Eng(e)$.

A \emph{Gibbs measure} on $\M(G)$ is a probability measure with the
following property.  If we fix the matching in an annular region, the matchings
inside and outside the annulus are independent of each other and
the probability of any interior matching $M$ is proportional to $e^{-\Eng(M)}$.
An \emph{ergodic Gibbs measure (EGM)} is a Gibbs measure on $\M(G)$ which is invariant and ergodic
under the action of $\Z^2$.

For an EGM $\mu$ let $s=\E[h(v+(1,0))-h(v)]$ and $t=\E[h(v+(0,1))-h(v)]$ be the expected horizontal
and vertical height change. We then have $\E[h(v+(x,y))-h(v)]=sx+ty$. We call $(s,t)$ the
\emph{slope} of $\mu$.

\subsubsection{Gibbs measures of fixed slope}
On $\M(G_n)$ we define a probability measure $\mu_n$ satisfying
$$
\mu_n(M) = \frac{e^{-\Eng(M)}}{Z}\,,
$$
for any matching $M\in \M(G_n)$. Here $Z$ is a normalizing constant
known as the \emph{partition function}.

For a fixed $(s,t)\in\R^2$, let $\M_{s,t}(G_n)$ be
the set of matchings of $G_n$ whose height change is
$(\lfloor ns\rfloor,\lfloor nt\rfloor)$.
Assuming that $\M_{s,t}(G_n)$ is nonempty, let
$\mu_n(s,t)$ denote the conditional measure induced by $\mu_n$
on $\M_{s,t}(G_n)$.

The following results are found in Chapters 8 and 9 of \cite{Sheffield}:

\begin{thm}[\cite{Sheffield}]\label{Sheff1} For each $(s,t)$ for which $\M_{s,t}(G_n)$ is nonempty
for $n$ sufficiently large, $\mu_n(s,t)$ converges as $n\to\infty$ to an EGM $\mu(s,t)$ of slope
$(s,t)$. Furthermore $\mu_n$ itself converges to $\mu(s_0,t_0)$ where $(s_0,t_0)$ is the limit of
the slopes of $\mu_n$. Finally, if $(s_0,t_0)$ lies in the interior of the set of $(s,t)$ for which
$\M_{s,t}(G_n)$ is nonempty for $n$ sufficiently large, then every EGM of slope $(s,t)$ is of the
form $\mu(s,t)$ for some $(s,t)$ as above; that is, $\mu(s,t)$ is the unique EGM of slope $(s,t)$.
\end{thm}

\subsubsection{Surface tension}\label{ssrft}

Let $$Z_{s,t}(G_n)=\sum_{M\in\M_{s,t}(G_n)}e^{-\Eng(M)}$$ be the partition function
of $\M_{s,t}(G_n)$.
Define $$Z_{s,t}(G)=\lim_{n\to\infty}Z_{s,t}(G_n)^{1/n^2} .$$ The existence of this limit is easily
proved using subadditivity as in \cite{CKP}.  The function $Z_{s,t}(G)$ is the \emph{partition
function per fundamental domain} of $\mu(s,t)$ and
$$
\sigma(s,t) = -\log Z_{s,t}(G)
$$
is called the
\emph{surface tension} or \emph{free energy} per fundamental domain.
The explicit form of this function is obtained in Theorem \ref{legendrethm}.

The measure $\mu(s_0,t_0)$ in Theorem \ref{Sheff1} above is the one which has
minimal free energy per fundamental domain.  Since the
surface tension is strictly convex (see Chapter 8 of \cite{Sheffield} or Theorem \ref{aboutsigma}
below), the surface-tension minimizing slope is unique and equal to $(s_0, t_0)$.

\subsection{Gauge equivalence and magnetic field}

\subsubsection{Gauge transformations}

Since $G$ is bipartite, each edge $e=(\ww,\bb)$ has a natural orientation:
from its white vertex $\ww$ to its black vertex $\bb$.
Any function $f$ on the edges can therefore be canonically identified with a
$1$-form, that is, a function on oriented  edges satisfying
$f(-e)=-f(e)$, where $-e$ is the edge $e$ with its opposite orientation.
We will denote by $\Omega^1(G_1)$ the linear space of $1$-forms on $G_1$.
Similarly, $\Omega^0$ and $\Omega^2$ will denote functions on
vertices and oriented faces, respectively.

The standard differentials
$$
0 \to \Omega^0 \xrightarrow{\,d\,}  \Omega^1 \xrightarrow{\,d\,}
\Omega^2 \to 0
$$
have the following concrete meaning in the dimer problem. Given two
energy functions $\Eng_1$ and $\Eng_2$, we say that they are
\emph{gauge equivalent} if
$$
\Eng_1 = \Eng_2 + df \,, \quad f\in \Omega^0\,,
$$
which means that for every edge $e=(\ww,\bb)$
$$
\Eng_1(e) = \Eng_2(e) + f(\bb)-f(\ww) \,,
$$
where $f$ is some function on the vertices. It is clear that
for any perfect matching $M$, the difference $\Eng_1(M)- \Eng_2(M)$
is a constant independent of $M$, hence the energies $\Eng_1$
and $\Eng_2$ induce the same probability distributions on
dimer configurations.

\subsubsection{Rotations along cycles}

Given an oriented cycle
$$
\gamma = \{\ww_0,\bb_0,\ww_1,\bb_1,\dots,\bb_{k-1},\ww_k\}\,,
\quad \ww_k=\ww_0\,,
$$
in the graph $G_1$, we define
$$
\int_\gamma \Eng = \sum_{i=1}^{k-1}
\big[\Eng(\ww_i,\bb_i)-\Eng(\ww_{i+1},\bb_i)\big]\,.
$$
It is clear that $\Eng_1$ and $\Eng_2$ are gauge
equivalent if and only if $\int_\gamma \Eng_1 = \int_\gamma \Eng_2$
for all cycles $\gamma$.  We call $\int_\gamma \Eng$ the
\emph{magnetic flux} through $\gamma$. It measures the
change in energy under the following basic transformation
of dimer configurations.

Suppose that a dimer configuration $M$ is such that
every other edge of a cycle $\gamma$ is included in $M$.
Then we can form a new configuration $M'$ by
$$
M' = M \, \triangle \, \gamma\,,
$$
where $\triangle$ denotes the symmetric difference.
This operation is called \emph{rotation along $\gamma$}. It is
clear that
$$
\Eng(M') = \Eng(M) \pm \int_\gamma \Eng \,.
$$
The union of any two perfect matchings $M_1$ and $M_2$ is
a collection of closed loops and one can obtain $M_2$ from
$M_1$ by rotating along all these loops. Therefore, the
magnetic fluxes uniquely determine the relative weights
of all dimer configurations.

\subsubsection{Magnetic field coordinates}\label{smag}

Since the graph $G_1$ is embedded in the torus,   the
gauge equivalence classes of energies are parametrized by
 $\R^{F-1}\oplus\R^2$, where $F$ is the number faces of $G_1$.
The first summand is $d\Eng$, a function on the faces
subject to one relation: the sum is zero. We will denote
the function $d\Eng\in \Omega^2(G_1)$ by $B_z$. The other two
parameters
$$
(B_x,B_y)\in \R^2
$$
are the magnetic flux along a cycle winding once horizontally
(resp.\ vertically) around the torus.

In practice we will fix $B_z$ and vary $B_x,B_y$, as follows.
Let $\gamma_x$ be a path in the dual of $G_1$ winding once horizontally
around the torus. Suppose that $k$ edges of $G_1$ are crossed by
$\gamma_x$. On each edge of $G_1$ crossed by $\gamma_x$,
add energy $\pm\frac1k\Delta B_x$ according to whether the upper vertex
(the one to the left of $\gamma_x$ when $\gamma_x$ is oriented in the
positive $x$-direction)
is black or white. Similarly, let $\gamma_y$ be a vertical path in the dual
of $G_1$, crossing $k'$ edges of $G_1$; add $\pm\frac1{k'}\Delta B_y$
to the energy of edges crossed by $\gamma_y$
according to whether the left vertex is black or white.

The new magnetic field is now $B'=B+(0,\Delta B_x,\Delta B_y)$.
This implies that the change in energy of a
matching under this change in magnetic field depends linearly on the
height change of the matching:

\begin{lem}\label{htcontribution} For a matching $M$ of $G_1$ with
height change $(h_x,h_y)$ we have
$$
\Eng_{B'}(M)-\Eng_{B'}(M_0)=\Eng_{B}(M)-\Eng_{B}(M_0)+\Delta B_x h_x+
\Delta B_y h_y\,.
$$
\end{lem}


\section{Surface tension}

\subsection{Kasteleyn matrix and characteristic polynomial}
\subsubsection{Kasteleyn weighting} A Kasteleyn matrix for a finite bipartite
\emph{planar} graph $\Gamma$ is a
weighted, signed adjacency matrix for $\Gamma$, whose determinant is the partition function for
matchings on $\Gamma$. It can be defined as follows. Multiply the edge weight of each edge of $\Gamma$ by $1$
or $-1$ in such a way that the following holds: around each face there are an odd number of $-$
signs if the face has $0\bmod 4$ edges, and an even number if the face has $2\bmod 4$ edges.  This
is always possible \cite{Kast1}.

Let $K=(K_{\ww\bb})$ be the matrix with rows indexing the
white vertices and columns indexing the black vertices,
with $K_{\ww\bb}$ being the above signed edge weight $\pm e^{-\Eng((\ww,\bb))}$
(and $0$ if there is no edge).
Kasteleyn proved \cite{Kast1} that $|\det K|$ is the partition function,
$$|\det K|=Z(\Gamma)=\sum_{m\in\M(\Gamma)}e^{-\Eng(m)}.$$

\subsubsection{Periodic boundary conditions}
For bipartite graphs on a torus $\T^2$, one can construct a Kasteleyn matrix $K$ as above
\cite{Kast1}.
Then $|\det K|$ is a signed sum of weights of matchings,
where the sign of a matching
depends on the parity of its horizontal and vertical height change.
This sign is a function on $H_1(\T^2,\Z/2)$,
that is, matchings with the same horizontal and vertical height change
modulo $2$ appear in $\det K$ with the same sign.
Moreover of the four possibly parity classes, three have the
same sign in $\det K$ and one has the opposite sign \cite{Tesler}.
In other words, this sign function is one of the 4
\emph{spin structures}
or theta characteristics on the torus.

Which one of the 4 spin structures
it is depends on
the choices in the definition of the Kasteleyn matrix.
By an appropriate choice we can make the $(0,0)$ parity class
(whose height changes are both even) have even sign and the remaining
classes have odd sign, that is, $\det K=M_{00}-M_{10}-M_{01}-M_{11},$
where $M_{00}$ is the partition function for matchings with even
horizontal and vertical height changes, and so on.

The actual partition function can then be obtained as a sum of four determinants
$$Z=\frac12(-Z^{(00)}+Z^{(10)}+Z^{(01)}+Z^{(11)}),$$ where $Z^{(\theta\tau)}$ is the determinant of
$K$ in which the signs along a horizontal dual cycle (edges crossing a horizontal path in the dual)
have been multiplied by $(-1)^{\theta}$ and along a vertical cycle have been multiplied by
$(-1)^\tau$. (Changing the signs along a horizontal dual cycle has the effect of negating the
weight of matchings with odd horizontal height change, and similarly for vertical.) For details see
\cite{Kast1, Tesler}.

\subsubsection{Characteristic polynomial} \label{sfund}
Let $K$ be a Kasteleyn matrix for the graph $G_1$ as above.
Given any positive parameters $z$ and $w$, we construct a ``magnetically altered'' Kastelyn matrix
$K(z,w)$ from $K$ as follows.

\begin{figure}[hbtp]\psset{unit=0.5 cm}
  \begin{center}
    \begin{pspicture}(-2,0)(12,10)
    \rput(5,5){\scalebox{0.64}{\includegraphics{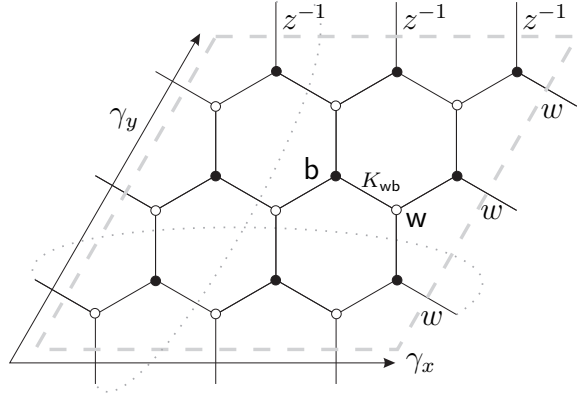}}}
    \rput[lb](4.7,9.4){$z^{-1}$}
    \rput[lb](7.9,9.4){$z^{-1}$}
    \rput[lb](11.1,9.4){$z^{-1}$}
    \rput[rt](9.0,1.9){$w$}
    \rput[rt](10.6,4.7){$w$}
    \rput[rt](12.2,7.4){$w$}
    \rput[rt](5.7,6){$\bb$}
    \rput[lt](8,4.5){$\ww$}
    \rput[lb](6.8,5.1){\scriptsize{$K_{\ww\bb}$}}
    \rput[l](8,0.5){$\gamma_x$}
    \rput[r](0.9,7){$\gamma_y$}
    \end{pspicture}
    \caption{\sl The operator $K(z,w)$}
    \label{fKzw}
  \end{center}
\end{figure}

Let $\gamma_x$ and $\gamma_y$ be the paths introduced
in Section \ref{smag}. Multiply each edge crossed by $\gamma_x$
by $z^{\pm 1}$ depending on whether the black vertex is
on the left or on the right, and similarly for $\gamma_y$.
See Figure \ref{fKzw} for an illustration of this
procedure in the case of the honeycomb graph with
$3\times 3$ fundamental domain.
We will refer to $P(z,w) = \det K(z,w)$ as the
\emph{characteristic polynomial} of $G$.

For example, for the square-octagon graph from Figure \ref{fsqoct} this
gives
$$
P(z,w) = z+\frac1z+w+\frac1w+5\,.
$$
Recall that $M_0$ denotes the
 reference matching  in the definition of the height function
and $\omega_0$ denotes the corresponding flow.  Let $x_0$ denote the
total flow of $\omega_0$ across $\gamma_x$ and similarly let $y_0$ the total flow of $\omega_0$
across $\gamma_y$.  The above remarks imply the following:

\begin{prop}\label{K=P} We have $$P(z,w)=z^{-x_0}w^{-y_0} \sum_{M\in \M(G_1)} e^{-\Eng(M)}z^{h_x
}w^{h_y}(-1)^{h_xh_y+h_x+h_y}$$ where $h_x=h_x(M)$ and $h_y=h_y(M)$ are the (integer)
horizontal and vertical height change of the matching $M$ and $\Eng(M)$ is its energy.
\end{prop}

Since $G_1$ has a finite number of matchings, $P(z,w)$ is a Laurent polynomial in $z$ and $w$ with
real coefficients. The coefficients are negative or zero except when $h_x$ and $h_y$ are both even.
Note that if the coefficients in the definition of $P$ were replaced with their absolute values
(i.e., if we ignored the $(-1)^{h_xh_y+h_x+h_y}$ factor), then $P(1,1)$ would be simply the
partition function $Z(G_1)$, and $P(z,w)$ (with $z$ and $w$ positive) would be the partition
function obtained using the modified energy $\Eng'(M) = \Eng(M) + h_x \log z + h_y \log w$.  With
the signs, however, $P((-1)^{\theta}, (-1)^{\tau}) = Z^{(\theta\tau)}$ and the partition function
may be expressed in terms of the characteristic polynomial as follows:
$$Z = \frac12 \left( - P(1,1) + P(1,-1) + P(-1,1) + P(-1,-1)\right)$$

As we will see, all large-scale properties of the dimer model
depends only on the polynomial $P(z,w)$.

\subsubsection{Newton polygon and allowed
slopes}
By definition,
the Newton polygon $N(P)$ of $P$ is the convex hull in $\R^2$ of the set of integer
exponents of monomials in $P$, that is
$$N(P)=\textup{convex hull}\left\{(j,k)\in\Z^2\,\big|\, z^jw^k \mbox{ is a
monomial in }P(z,w)\right\}.$$
\begin{prop}
The Newton polygon is the set of possible slopes of EGMs,
that is, there exists an EGM $\mu(s,t)$ if and only if $(s,t)\in N(P)$.
\end{prop}

\begin{proof}
First we prove that $(s,t)$ must lie in $N(P)$. A translation-invariant
measure of average slope $(s,t)$ determines a unit white-to-black
flow on $G_1$ with vertical flux
$s$ and horizontal flux $t$: the flow along an edge is the probability
of that edge occurring.
However the set of matchings of $G_1$ are the vertices of the polytope of
unit white-to-black flows of $G_1$, and the height change $(s,t)$ is a
linear function on this polytope. Therefore $(s,t)$ is contained in $N(P)$.

The absolute value of the coefficient of $z^iw^j$ in $P$ is the weighted sum of
matchings of $G_1$ with height change $(i,j)$. Therefore there is a matching corresponding to each
slope at the corners of $N(P)$. Next, we show that given two periodic matchings of slopes
$(s_1,t_1)$ and $(s_2,t_2)$, there is a periodic matching (possibly with larger period) with their
average slope. Suppose the two matchings have the same period (otherwise take the lcm of their
periods). The union of the two matchings is a set of cycles; one can change from one matching to
the other by rotating along each cycle. If the slopes $(s_1,t_1)$ and $(s_2,t_2)$ are
unequal some of these cycles necessarily have nonzero homology the torus, so that rotating along
them will change the slope. On a fundamental domain of twice the size, shift one of each pair of
all such cycles; this way one creates a matching with slope which is the average of the two initial
slopes.  Taking limits of these periodic constructions, one can produce a tiling with asymptotic
slope corresponding to any $w \in N(P)$. \end{proof}

Note that changing the reference matching $M_0$ in the definition of the height function merely
translates the Newton polygon.

\subsection{Asymptotics}

\subsubsection{Enlarging the fundamental domain}

Characteristic polynomials of larger graphs may be computed recursively as follows:

\begin{thm} \label{Precursive} Let $P_n$ be the characteristic polynomial of $G_n$.  Then $$P_n(z,w) =
\prod_{z_0^n = z,} \prod_{w_0^n = w} P(z_0,w_0).$$ \end{thm}
\begin{proof}
We follow the argument of  \cite{CKP} where this
fact is proved for grid
graphs.  Since
symmetry implies that the right side is a polynomial in $z$ and $w$, it is enough to check this
statement for positive values of $z$ and $w$. View the Kastelyn matrix $K_n(z,w)$ of $G_n$ as a
linear map from the space $V_w$ of functions on white vertices of $G_n$ to the space $V_b$ of
functions on black vertices. When $\alpha$ and $\beta$ are $n$th roots of unity, let $V_w^{\alpha,
\beta}$ and $V_b^{\alpha, \beta}$ be the subspaces of functions for which translation by one period
in the horizontal or vertical direction corresponds to multiplication by $\alpha$ and $\beta$
respectively.  Clearly, these subspaces give orthogonal decompositions of $V_w$ and $V_b$, and
$K_n(z,w)$ is block diagonal in the sense that it sends an element in $V_w^{\alpha,\beta}$ to an
element in $V_b^{\alpha, \beta}$. We may thus write $\det K_n(z,w)$ as a product of the
determinants of the $n^2$ restricted linear maps from $V_w^{\alpha, \beta}$ to $V_b^{\alpha,
\beta}$; these determinants are given by $\det K (\alpha z^{1/n},\beta w^{1/n})$. \end{proof}

This recurrence relation allows us to compute partition functions on general $G_n$ in terms of $P$:
\begin{cor} \begin{equation}\label{Z4}
Z(G_n)=\frac12(-Z_n^{(00)}+Z_n^{(01)}+Z_n^{(10)}+Z_n^{(11)}), \end{equation} where
\begin{equation}\label{prodform} Z_n^{(\theta\tau)}=P_n( (-1)^\theta, (-1)^{\tau}) =
\prod_{z^n=(-1)^\theta,}\prod_{w^n=(-1)^\tau} P(z,w). \end{equation} \end{cor}

\subsubsection{Partition function per fundamental domain}

We are interested in the asymptotics of $Z(G_n)$ when $n$ is large. The
logarithm of the expression (\ref{prodform}) is a Riemann sum for an integral over the unit torus
$\T^2=\{(z,w)\in\C^2~:~|z|=|w|=1\}$ of $\log P$; the evaluation of this is a Riemann sum so we have
$$\frac1{n^2}\log Z_n^{(\theta\tau)}= \frac1{(2\pi i)^2}\int_{\T^2}\log P(z,w)\frac{dz}z\frac{dw}w
+ o(1)$$ on condition that none of the points
\begin{equation}
  \label{sumpt}
  \{(z,w):z^n=(-1)^\theta,w^n=(-1)^\tau\}
\end{equation}
falls close to a zero of $P$. If it does, such a point will affect
the sum only if it falls within $e^{-O(n^2)}$ of a zero of
$P$ (and in any case can only decrease the sum). In this case, for any
$n'$ near but
not equal to $n$, no point of the form \eqref{sumpt} will fall so close
to this zero of $P$. In Theorem
\ref{maximal} below we prove that $P$ has at most two simple zeros on the unit torus. It follows that only for a very rare set of $n$ does
the Riemann sum not approximate the
integral.

Since the partition function $Z(G_n)$ satisfies $$Z_n^{(\theta\tau)}\leq Z(G_n) \leq
2\max_{\theta,\tau}\{Z_n^{( \theta,\tau)}\},$$ we have $${\lim_{n\to\infty}}'\frac1{n^2}\log
Z(G_n)= \frac1{(2\pi i)^2}\int_{\T^2}\log P(z,w)\frac{dz}z\frac{dw}w,$$ where the $\lim'$ means
that the limit holds except possibly for a rare set of $n$s. But now a standard subadditivity
argument (see e.g. \cite{CKP}) shows that $Z(G_n)^{1/n^2}\leq Z(G_m)^{1/m^2}(1+o(1))$ for all large
$m$ so that in fact the limit exists without having to take a subsequence.

\begin{thm} \label{logZexpression} Under the assumption that $P(z,w)$ has a finite number of zeros
on the unit torus $\T^2$, we have $$\log Z\stackrel{\rm{def}}{=}\lim_{n\to\infty}\frac1{n^2}\log
Z(G_n)= \frac1{(2\pi i)^2}\int_{\T^2}\log P(z,w)\frac{dz}z\frac{dw}w.$$ \end{thm}

The quantity $Z$ is the partition function per fundamental domain.

\subsubsection{The amoeba and Ronkin function of a polynomial}

Given a polynomial $P(z,w)$, its \emph{Ronkin function} is by
definition the following integral
\begin{equation}
  \label{Ronkin}
  F(x,y) =
\frac1{(2\pi i)^2}\int_{\T^2} \log P(e^{x}z,e^{y}w)\frac{dz}z\frac{dw}w \,.
\end{equation}
A closely related object is the \emph{amoeba} of the polynomial $P$
defined as the image of the curve $P(z,w)=0$ in $\C^2$
under the map
$$
(z,w)\mapsto(\log|z|,\log|w|) \,.
$$
We will call the curve $P(z,w)=0$ the \emph{spectral
curve} and denote its amoeba  by $\A(P)$.

\begin{figure}[!hbtp]
  \begin{center}
    \scalebox{0.5}{\includegraphics{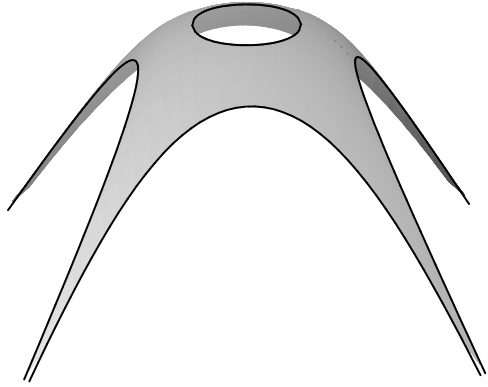}}
    \caption{\sl The curved part of minus the
Ronkin function of $z+\frac1z+w+\frac1w+5$. This is the
limit height function shape for square-octagon dimers
with crystal corner boundary conditions.}
    \label{fRonk}
  \end{center}
\end{figure}

\begin{figure}[!hbtp]
  \begin{center}
    \scalebox{1}{\leavevmode \epsfbox[95 431 364 721]{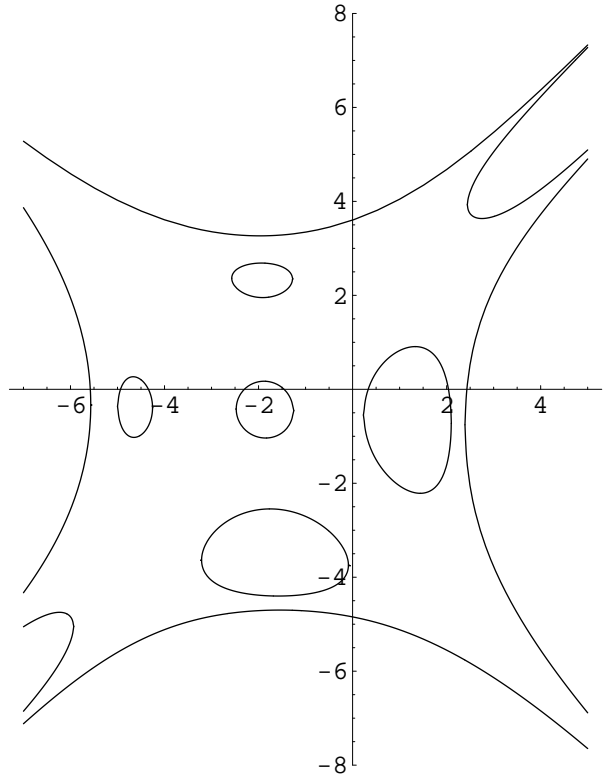}}
    \caption{
\sl The amoeba corresponding to the model of Figure \ref{twophases}.  
Its complement has one bounded component for each of the five interior integer
points of $N(P) = \{x: |x| \leq 2\}$. It also has two ``semi-bounded'' (i.e., contained in a
strip of finite width) components, corresponding to two of the non-corner integer points on the
boundary of $N(P)$.  The four large components correspond to the corner vertices of $N(P)$.  In the case of equal weights, the holes in the 
amoeba shrink to points
and only four large unbounded components of the complement are
present. 
}
    \label{am4by4}
  \end{center}
\end{figure}

It is clear that
the integral \eqref{Ronkin} is singular if and only if
$(x,y)$ lies in the amoeba. In fact, the Ronkin function
is linear on each component of the amoeba complement and
strictly convex over the interior of the amoeba (in
particular, implying that each component of $\R^2\setminus \A(P)$
is convex).
This and many other useful facts about the amoebas and
Ronkin function can be found in \cite{Mikhalkin}. See Figures
\ref{fRonk} and \ref{am4by4} for an illustration of these
notions.


We distinguish between
the unbounded complementary components and
the bounded complementary components.


\subsubsection{Surface tension}\label{ssurft}

Theorem \ref{Sheff1} gave, for a fixed magnetic field,
a two-parameter family of EGMs $\{\mu(s,t)\}$. Let us vary the
magnetic field as in Section \ref{smag}.
Let $\mu_n(B_x,B_y)$ be the measure on dimer configurations
on $G_n$ in the presence of an additional parallel magnetic field
$B_x,B_y$. We define $\mu(B_x,B_y)$ to be the limit of $\mu_n(B_x,B_y)$
as $n\to\infty$ and let $Z_{B_x,B_y}$ be its partition function
per fundamental domain.

We can compute $Z_{B_x,B_y}$
in two different ways. On the one hand,
the characteristic polynomial becomes $P(e^{B_x}z,e^{B_y}w)$
and, hence, by Theorem \ref{Precursive} we have $Z_{B_x,B_y}=F(B_x,B_y)$,
where $F$ is the the Ronkin function of $P$. On the other
hand, using Lemma \ref{htcontribution} and basic properties of
the surface tension (see Section \ref{ssrft}) we obtain
\begin{equation}
\label{ldeq} F(B_x,B_y) = \max_{(s,t)} \left(- \sigma(s,t)+s B_x + t B_y \right)\,.
 \end{equation}
In other words, $F$ is the Legendre dual of the surface tension. Since
the surface tension is strictly convex, the Legendre transform is
involutive and we obtain the following

\begin{thm}\label{legendrethm} The surface tension $\sigma(s,t)$ is
the Legendre transform of the Ronkin function of the
characteristic polynomial $P$.
\end{thm}

Recall that the Ronkin function is linear on each component of
the amoeba complement. We will call the corresponding flat
pieces of the graph of the Ronkin function \emph{facets}.
They correspond to conical singularities (commonly referred
to as ``cusps'') of the surface tension $\sigma$. The gradient
of the Ronkin function maps $\R^2$ to the Newton polygon $N(P)$.
It is known that the slopes of the facets form a subset of
the integer points inside $N(P)$ \cite{Mikhalkin}. Therefore, we have the following immediate
corollary:

\begin{cor} \label{aboutsigma} The surface tension $\sigma$ is strictly
convex and is smooth on the interior of $N(P)$, except at a subset of
points in $\mathbb Z^2 \cap
N(P)$. Also, $\sigma$ is a piecewise linear function on $\partial N(P)$, with no slope
discontinuities except at a subset of points in $\mathbb Z^2 \cap \partial N(P)$.
\end{cor}

In Section \ref{smaxim}, we will see that the spectral curves of
dimer models are always very special real plane curves. As a
result, their amoebas and Ronkin function have a number
of additional nice properties, many of which admit a
concrete probabilistic interpretation.

The following Figure \ref{fsurf} shows the Legendre dual
of the Ronkin function from Figure \ref{fRonk}. It is
the surface tension function for certain periodically
weighted dimers on the square grid with $2\times 2$
fundamental domain and also for the uniformly weighted
dimers on the square-octagon graph.
\begin{figure}[!hbtp]
  \begin{center}
    \scalebox{0.5}{\includegraphics{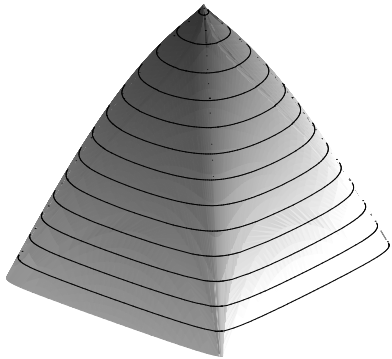}}
    \caption{\sl (negative of) Surface tension for the square-octagon graph}
    \label{fsurf}
  \end{center}
\end{figure}

\section{Phases of the dimer model}\label{sphas}

\subsection{Frozen, liquid, and gaseous phases}

We will show that EGMs with distinctly different
qualitative properties are possible in a general
periodic dimer model. The different types of
behavior can be classified as frozen, liquid, and gaseous.
We will take the fluctuation of the height
function as the basis for the classification.
This, as it will turn out, is equivalent to
the classification by the rate of decay of correlations.

Let $f$ and $f'$ be two faces of the graph $G$ and
consider the height function difference $h(f) - h(f')$.
An EGM is called a \emph{frozen
phase} if some of the height differences are deterministic---i.e., there exist distinct $f$ and
$f'$ arbitrarily far apart for which $h(f) - h(f')$ is deterministic.
An example is the delta-measure on the
brick-wall matching of the square grid.

\begin{figure}[!hbtp]
  \begin{center}
    \scalebox{0.55}{\leavevmode \epsfbox[50 50 450 450]{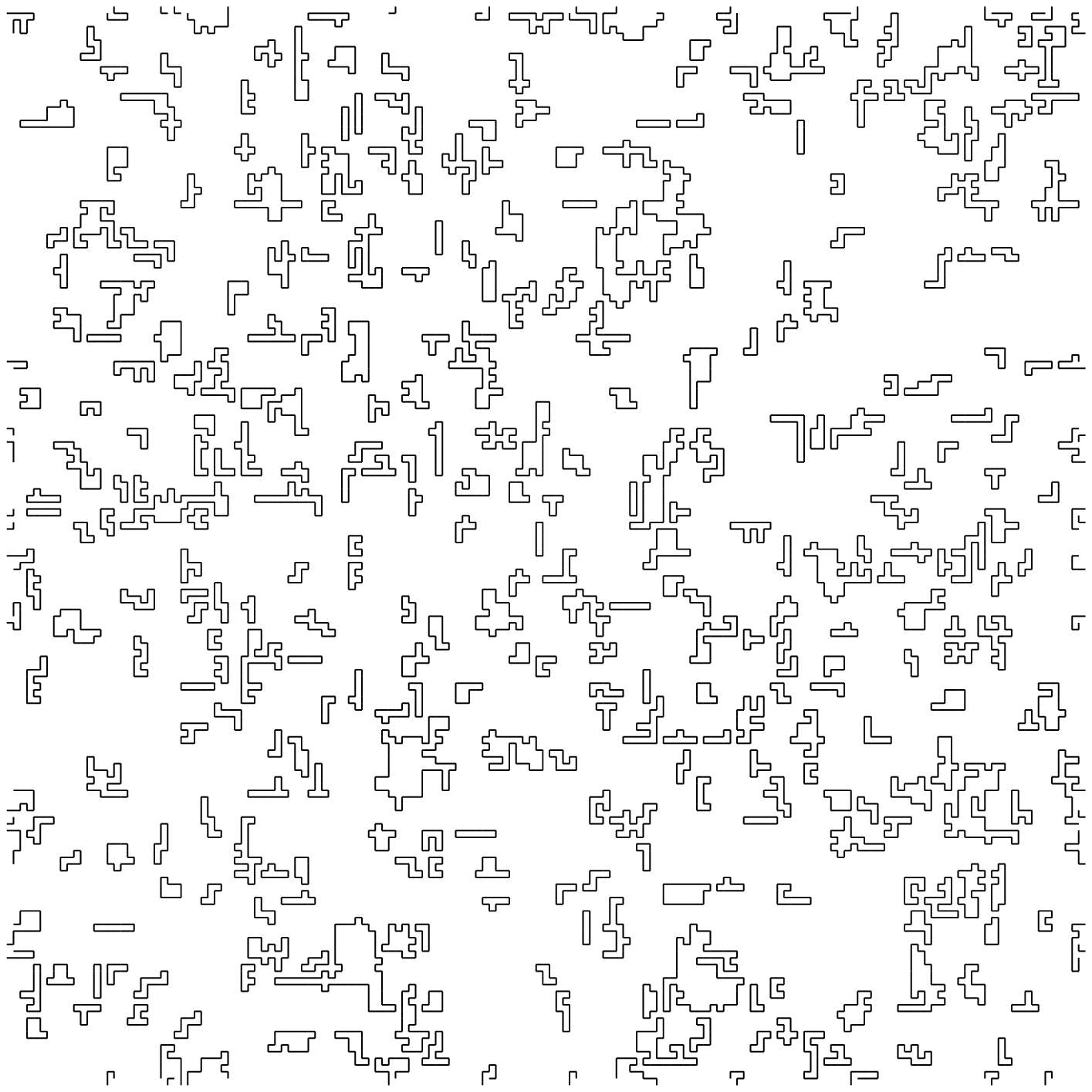}} 
\hfill
\scalebox{0.55}{\leavevmode \epsfbox[50 50 450 450]{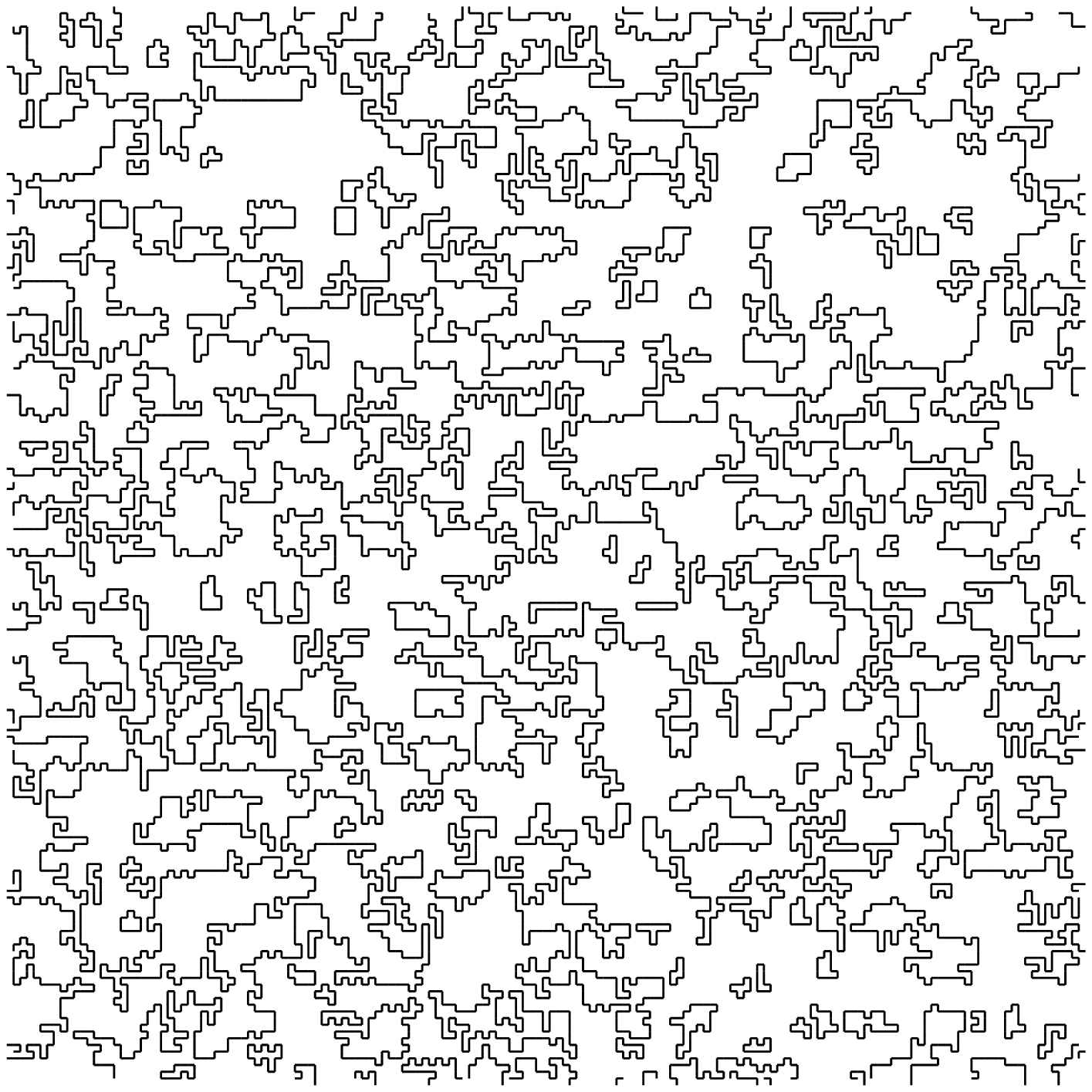}}
\caption{\sl All cycles of length ten or longer in the union of two random perfect
matching of $\mathbb Z^2$ with $4\times 4$ fundamental domain. The weight
of one edge equals to $10$ and all other edges have weight $1$. The amoeba
for this case is plotted in Figure \ref{am4by4}. The slope
on the left is $(0,0)$ and this is a smooth phase. The rough phase on the
right has slope $(0,0.5)$.}
    \label{twophases}
  \end{center}
\end{figure}

A non-frozen EGM $\mu$ is called a \emph{gaseous phase} or \emph{smooth phase} if the height
fluctuations have bounded variance, i.e., the $\mu$ variance of the random variable $h(f) -
h(f')$ is bounded independently of $f$ and $f'$.  A non-frozen EGM $\mu$ is called a \emph{liquid
phase} or \emph{rough phase} if the $\mu$-variance of the height difference is not
bounded. The difference between the smooth and rough phases is
illustrated in Figure \ref{twophases}. 

We will prove in Theorem \ref{variance} that
in the liquid phase the variance of $h(f) -h(f')$ grows \emph{universally} like
$\pi^{-1}$ times the logarithm of the distance between $f$ and $f'$.
The following is our main result about phases:

\begin{thm}\label{mainmsr} The measure $\mu(B_x,B_y)$ is frozen, liquid, or gaseous, respectively,
when $(B_x,B_y)$ is respectively in the closure of an unbounded complementary component of $\A(P)$,
in the interior of $\A(P)$, or in the closure of a bounded complementary component of $\A(P)$.
\end{thm}

This theorem is proved in the next three sections.
In Corollary 9.1.2 of \cite{Sheffield} there is a different proof
that when $(s,t)$ lies in the interior of $N(P)$,
$\mu(s,t)$ can only be smooth if $s$ and $t$ are integers.

We will see that in the liquid and gaseous phases the edge-edge correlations
decay polynomially and exponentially, respectively. In the frozen case,
some edge-edge correlations do not decay at all.



\subsection{Frozen phases}
\subsubsection{Matchings and flows}
Recall the interpretation of a matching as a black-to-white unit flow.
If $M$ is a matching and $M_0$ is the reference matching in the definition
of the height function, then the difference of the flows $M-M_0$
defines a divergence-free flow.
The height function of $M$ is the corresponding flux, that is,
for two faces $f_1,f_2$,
$h(f_2)-h(f_1)$ is the
amount of flow crossing any dual path from $f_1$ to $f_2$.
For two adjacent faces $f_1,f_2$
let $d(f_1,f_2)$ be the maximal possible (oriented) flow along the
edge $e$ between them (where $e$ is oriented so that $f_1$ is on its left).
This is the forward capacity of the oriented edge $e$.
That is, if $e\not\in M_0$, its capacity
is $1$ from its black vertex to its white vertex,
and $0$ in the other direction;
if $e\in M_0$, its capacity is $1$ from its white vertex to its black vertex,
and zero in the reverse direction.
For any two faces $f_1$ and $f_2$
let $D(f_1,f_2)$ be the minimum, over all dual paths from $f_1$ to $f_2$,
of the sum of the capacities
of the segments oriented to cross the path from left to right. By the max-flow-min-cut theorem,
a function $h$ is the height function for a tiling if and only if
$$\forall f_1,f_2\hskip1cm D(f_1,f_2)\geq h(f_2)-h(f_1).$$
See \cite{Thurston, Fournier} for a reference.

Now let $(s,t)\in\R^2$. If there is no tiling with height function having slope $(s,t)$
then there is a face $f$ and  $(x,y)\in\Z^2$ such that  $D(f,f+(x,y))< sx+ty$.
We claim that in this case there is a face path from $f$
to some translate $f+(x',y')$ on which $D(f,f+(x',y'))<sx'+ty'$
and all faces along this path are of distinct types, that is, are not translates of each other
(except for the first and last faces).
To see this, note that
if a face path $f_1,f_2,\dots,f_k$ passes through two faces of the same type,
say $f_i$ and $f_j$, then one of the two paths
$f_1,\dots,f_i,f_{j+1},\dots,f_k$ and $f_i,\dots,f_j$ will necessarily satisfy the
strict inequality.

But up to translation there are only a finite number of face paths
which start and end at the same face type
and which pass through each face type at most once.
Each such path gives one restriction on the slope:
$D(f_1,f_2)\geq sx+ty$ where $(x,y)$ is the translation of the path.

In particular the Newton polygon $N(P)$
is the set of $(s,t)$ defined by the intersection of the inequalities
$\{(s,t)~|~sx+ty\leq D(f_1,f_2)\}$, one for each of the above finite number of paths.
If $(s,t)$ is on the edge of $N(P)$, the path $\gamma$ from the corresponding
inequality has maximal flow, that is, in a tiling of slope $(s,t)$ all edges
on $\gamma$ are determined: they must occur with probability $1$ or $0$.

\subsubsection{Frozen paths}
When $(B_x,B_y)$ is in an unbounded component of the complement
of the amoeba, we prove that $\mu(B_x,B_y)$ is in a frozen phase.

The slope $(s,t)$ of $\mu(B_x,B_y)$ is an
integer point on the boundary of $N(P)$.
By the argument of the previous section, there is a face path $\gamma$ on
$G_1$, with homology class perpendicular to
$(s,t)$,  for which every edge crossing each lift of $\gamma$ is present with probability
$1$ or $0$ for $\mu(B_x,B_y)$. These lifts constitute \emph{frozen paths}
in the dual $G'$.
Edges which are in different components of the complement of the set of frozen
paths are independent.

For each corner of $N(P)$ there are two sets of frozen paths,  with different asymptotic directions.
The components of the complements of these paths are finite sets of edges.
The edges in each set are independent of all their translates.

\subsection{Edge-edge correlations}
For a finite planar graph $\Gamma$ the inverse of the Kasteleyn matrix
determines the edge probabilities: the probability of a set of edges
$\{e_1,\dots,e_k\}$ being in a random matching is the determinant
of the corresponding submatrix of $K^{-1}$, times the product of
the edge weights \cite{Kenyon}.

For a graph on a torus the corresponding statement is more complicated: We have
\begin{thm}[\cite{CKP}] The probability of edges $\{e_1=(\ww_1,\bb_1),\dots,e_k=(\ww_k,\bb_k)\}$
occurring is in a random matching of $G_n$ equals $\prod K(\ww_j,\bb_j)$ times
\begin{multline}\label{K4inv}
\frac12\left(\frac{-Z^{(00)}_n}{Z}\det(K_{00}^{-1}(\bb_j,\ww_i))+
\frac{Z^{(10)}_n}{Z}\det(K_{10}^{-1}(\bb_j,\ww_i))\right.\\
+\left. \frac{Z^{(01)}_n}{Z}\det(K_{01}^{-1}(\bb_j,\ww_i))+
\frac{Z^{(11)}_n}{Z}\det(K_{11}^{-1}(\bb_j,\ww_i))\right)
\end{multline}
\end{thm}
Here the determinants $\det(K_{\theta\tau}^{-1}(\bb_j,\ww_i))$ are $k\times k$ minors of
$K_{\theta\tau}^{-1}$. The asymptotics of this expression are again complicated by the zeros of $P$
on $\T^2$. The entries in $K_{\theta\tau}^{-1}$ have the form (see \cite{CKP})
$$
K_{\theta\tau}^{-1}(\bb,\ww)= \frac1{n^2}\sum_{z^n=(-1)^\theta}\sum_{w^n=(-1)^\tau}
\frac{Q(z,w)\,w^xz^y}{P(z,w)}
$$
where $Q(z,w)$ is one of a finite number of polynomials (depending on
where $\ww$ and $\bb$ sit in their respective fundamental domains:
$Q/P$ is an entry of $K(z,w)^{-1}$
where $K(z,w)$ is the matrix of Lemma \ref{K=P}) and $(x,y)\in\Z^2$ is the translation taking the
fundamental domain containing $\ww$ to the fundamental domain containing $\bb$.

This expression is a Riemann sum for the integral $$\frac1{(2\pi i)^2}\int_{\T^2}
\frac{Q(z,w)w^xz^y}{P(z,w)}\frac{dw}w\frac{dz}z,$$ except near the zeros of $P$. However the
contribution for the root $(z,w)$ nearest to a zero of $P$ is negligible unless $(z,w)$ is at
distance $O(\frac1{n^2})$ of the zero. But if this is the case then replacing $n$ with any $n'$ at
distance at least $O(\sqrt{n})$ from $n$ makes the contribution for this root negligible. Thus we
have
$$\lim_{n\to\infty}\!\!{}'\,
K_{n,\theta\tau}^{-1}(\ww,\bb)= \frac1{(2\pi i)^2}\int_{\T^2}
\frac{Q(z,w)w^xz^y}{P(z,w)}\frac{dw}w\frac{dz}z\,,
$$
where the limit is taken along a subsequence of
$n$'s.

Since all the $K^{-1}_{n,\theta\tau}$ have the same limit along a subsequence of $n$s, their
weighted average (as in (\ref{K4inv})) with weights $\pm Z_{\theta\tau}/2Z$ (weights which sum to
one and are bounded between $-1$ and $1$) has the same (subsequential) limit. This subsequential
limit defines a Gibbs measure on $\M(G)$. By Theorem \ref{Sheff1}, this measure is the \emph{unique}
limit of the Boltzmann measures on $G_n$. Thus we have proved
\begin{thm}
\label{limitingGibbs} For
the limiting Gibbs measure $\mu=\lim_{n\to\infty} \mu_n$, the probability of edges
$\{e_1,\dots,e_\ell\}$ where $e_j=(\ww_j,\bb_j)$, is $$\left(\prod_{j=1}^\ell K(\ww_j,\bb_j)\right)
\det(K^{-1}(\bb_k,\ww_j))_{1\leq j,k\leq \ell}\,,
$$
where, assuming $\bb$ and $\ww$ are in a single
fundamental domain,
\begin{equation}\label{Kinvint} K^{-1}(\bb,\ww+(x,y))=\frac1{(2\pi
i)^2}\int_{\T^2} K^{-1}(z,w)_{\bb\ww}\, w^xz^y\frac{dw}w\frac{dz}z.
\end{equation}
\end{thm}

We reiterate that $K^{-1}(z,w)_{\bb\ww}=Q_{\bb\ww}(z,w)/P(z,w),$
where $Q_{\bb\ww}$ is a polynomial in $z$ and $w$.

\subsection{Liquid phases (rough non-frozen phases)} \label{liquidsubsection}

\subsubsection{Generic case}

When $(B_x,B_y)$ is in the interior of the amoeba, $P(e^{B_x}z,e^{B_y}w)$
either has two simple zeros on the unit torus or a real node on the unit torus
 (see Theorem \ref{maximal}, below).
In the case of simple zeros (see Lemma \ref{fouriercoeffs} below),
$K^{-1}(\bb,\ww)$ decays
linearly but not faster, as $|w-b|\to\infty$. This implies that
the edge covariances decay quadratically:
\begin{multline}\notag
  \Cov(e_1,e_2):=\Pr(e_1\text{ and }e_2)-\Pr(e_1)\Pr(e_2)\\
= -K(\ww_1,\bb_1)K(\ww_2,\bb_2)K^{-1}(\bb_2,\ww_1)K^{-1}(\bb_1,\ww_2)\,.
\end{multline}
In section \ref{realnode} we show that in the case of a real node we have similar behavior.

\begin{lem}\label{fouriercoeffs} Suppose that $|z_0|=|w_0|=1$,
$\Im(-\beta w_0/\alpha z_0)>0,$ $x,y\in\Z$, and $R(z,w)$ is a smooth function on $\T^2$ with a
single zero at $(z_0,w_0)$ and satisfying
$$R(z,w)=\alpha(z-z_0)+\beta(w-w_0)+O(|z-z_0|^2+|w-w_0|^2).$$ Then we have the
following asymptotic formula for the Fourier coefficients of $R^{-1}$:
$$
\frac1{(2\pi i)^2}
\int_{\T^2}\frac{w^xz^y}{R(z,w)}\frac{dz}{z}\frac{dw}{w}= \frac{-w_0^xz_0^y}{2\pi i(x\alpha
z_0-y\beta w_0)}+O\left(\frac1{x^2+y^2}\right)\,.
$$ \end{lem}

Note that if $R$ has $k$ simple zeros, $1/R$ can be written as a sum of $k$ terms,
each of which is of the above form.

\begin{proof}
Replacing $z$ with $e^{i a}z_0$ and $w$ with $e^{i b}w_0$ we have
$$\alpha(z-z_0)+\beta(w-w_0)+O(\dots)=\alpha z_0 i a+\beta w_0 i b+O(\dots).$$
By adding a smooth function (whose Fourier coefficients decay at least quadratically)
to $1/R$ we can replace $1/R$
with
$$\frac{1}{\alpha z_0 i a+\beta w_0 i b}.$$
The integral is therefore
$$\frac{w_0^x z_0^y}{(2\pi)^2i}\int_{-\pi}^{\pi}\int_{-\pi}^{\pi}
\frac{e^{i(xb+ya)}}{\beta w_0 b+\alpha z_0 a}dadb+O(\dots)=
\frac{w_0^x z_0^y}{(2\pi )^2i}\int_{-\infty}^{\infty}\int_{-\infty}^{\infty}
\frac{e^{i(xb+ya)}}{\beta w_0 b+\alpha z_0 a}dadb+ O(\dots).$$
We first  integrate over the variable $a$: the integrand is a
meromorphic function of $a$ with a simple pole in the upper half plane if
$\Im(-\beta w_0b/\alpha z_0)>0$. Change the path of integration
to a path from $-N$ to $N$ followed by the upper half of a semicircle
centered at the origin of radius $N$.
The residue theorem then yields
$$\frac{w_0^x z_0^y}{2\pi \alpha z_0}\int_{0}^{\infty}e^{i(x-y\beta w_0/\alpha z_0)b}db
=\frac{w_0^x z_0^y}{2\pi \alpha z_0}\frac{-1}{i(x-y\beta w_0/\alpha z_0)}=
\frac{-w_0^x z_0^y}{2\pi i(x\alpha z_0-y\beta w_0)}$$
which gives the result.
\end{proof}

We now compute the variance in the height function.

\begin{thm}\label{variance} Suppose that the zeros of $P$ on $\T^2$ are
simple zeros at $(z_0,w_0)$ and $(\bar z_0,\bar w_0)$.
Let $\alpha,\beta$ be the derivatives of $P(z,w)$ with respect to $z$
and $w$ at $(z_0,w_0)$.
Then the height variance between two faces $f_1$ and $f_2$ is
$$\mbox{Var}[h(f_1)-h(f_2)]=\frac{1}{\pi}\log|\phi(f_1)-\phi(f_2)| +
o(\log|\phi(f_1)-\phi(f_2)| ),$$
where $\phi$ is the linear mapping $\phi(x+iy)= x\alpha z_0-y\beta w_0$.
\end{thm}

Since $\phi$ is a non-degenerate linear mapping, the above expression
for the variance is equivalent to
$\frac1{\pi}\log|f_1-f_2|+o(\log|f_1-f_2|).$ However it appears that
a slightly finer analysis would improve the little-o error in the statement to $o(1)$,
so we chose to leave the expression in the given form.
\medskip

\begin{proof}
Define $\tilde h=h-\E(h)$.
Let $f_1,f_2,f_3,f_4$ be four faces, all of which are far apart from each other.
We shall approximate $(\tilde h(f_1)-\tilde h(f_2))(\tilde h(f_3)-
\tilde h(f_4)).$
To simplify the computation we assume that $f_1$ and $f_2$ are translates
of each other, as well as $f_3$ and $f_4$.
Let $f_1=g_1,g_2,\dots,g_k=f_2$ be a path of translates of $f_1$ from $f_1$ to $f_2$,
with $g_{p+1}-g_p$ being a single step in $\Z^2$.
Similarly let $f_3=g_1',g_2',\dots,g_\ell'=f_4$ be a path from $f_3$ to $f_4$. We assume that
these paths are far apart from each other.

Then
\begin{equation}\label{varsum}
(\tilde h(f_1)-\tilde h(f_2))(\tilde h(f_3)-\tilde h(f_4))=
\sum_{p=1}^{k-1}\sum_{q=1}^{\ell-1}
 (\tilde h(g_{p+1})-\tilde h(g_p))(\tilde h(g_{q+1}')-h(g_q')).
\end{equation}
We consider one element of this sum at a time.
There are three cases to consider: when $g_{p+1}-g_p$ and $g'_{q+1}-g'_q$ are both horizontal, both vertical, and one vertical, one horizontal.

{}From Lemma \ref{fouriercoeffs}, for $x,y$ large we have
$$\int_{\T^2}\frac{Q(z,w)w^xz^y}{P(z,w)}\frac{dz}{z}\frac{dw}w=
-2\Re\left(\frac{w_0^xz_0^yQ(z_0,w_0)}{2\pi i(x\alpha z_0-y\beta w_0)}\right) +
O(\frac{1}{x^2+y^2}). $$

Recall that the matrix $Q(z,w)$ satisfies $Q(z,w)K(z,w)=P(z,w)\cdot\textup{Id}$.
Since $(z_0,w_0)$ is a simple zero of $P$, $K(z_0,w_0)$ has co-rank $1$.
In particular $Q(z_0,w_0)$ must have rank $1$. We write $Q(z_0,w_0)=UV^t$
where $V^tK(z_0,w_0)=0=K(z_0,w_0)U.$

Let $a_i=(\ww_i,\bb_i)$ be the edges crossing a ``positive" face path $\gamma$
from $g_p$ to $g_{p+1}$,
that is, a face path with the property that each edge crossed has its white vertex on the left.
Similarly
let $a'_j=(\ww_j',\bb_j')$ be the edges crossing a positive face path $\gamma'$
from $g_q'$ to $g_{q+1}'$.
Then
\begin{eqnarray*}
\E((\tilde h(g_{p+1})-\tilde h(g_p))(\tilde h(g_{q+1}')-\tilde h(g'_q)))&=&
\E(\sum_{i,j}(a_i-\bar a_i)(a_j'-\bar a_j'))\\&=&
\sum_{i,j}\E(a_ia_j')-\E(a_i)\E(a_j')\\
&=&-\sum_{i,j}K(\ww_i,\bb_i)K(\ww_j,\bb_j)K^{-1}(b'_j,\ww_i)K^{-1}(\bb_i,w'_j). \end{eqnarray*} Assuming these
faces $g_p,g_q'$ are far apart, this is equal to
\begin{multline}
  \notag
-\frac1{4\pi^2}\sum_{i,j}K(\ww_i,\bb_i)K(\ww_j,\bb_j) \times \\
\left(\frac{w_0^xz_0^y U_iV'_j}{
x\alpha z_0-y\beta w_0}+\frac{\bar w_0^x\bar z_0^y \bar U_i\bar V'_j}{ x\bar\alpha \bar
z_0-y\bar\beta \bar w_0}+O\left(\frac{1}{x^2+y^2}\right)\right) \times \\
\left(\frac{w_0^{-x}z_0^{-y}
U'_jV_i}{ x\alpha z_0-y\beta w_0}+\frac{\bar w_0^{-x}\bar z_0^{-y} \bar U'_j\bar V_i}{ x\bar\alpha
\bar z_0-y\bar\beta \bar w_0}+O\left(\frac{1}{x^2+y^2}\right)\right) \,.
\end{multline}

When we combine the cross terms we get an oscillating factor $z_0^{2x}w_0^{2y}$ or its
conjugate, which causes the sum (\ref{varsum}) of these terms when we sum over $p$ and $q$
to remain small.
So the leading term for fixed $p,q$ is
$$-\frac{2}{4\pi^2}\Re\left(\frac{1}{(x\alpha z_0-y\beta w_0)^2}\sum_{i,j}K(\ww_i,\bb_i)K(\ww_j,\bb_j)U_iV_iU_j'V'_j\right)$$
\begin{equation}\label{cross}=-\frac1{2\pi^2}\Re\left(\frac{1}{(x\alpha z_0-y\beta w_0)^2}(\sum_{i}K(\ww_i,\bb_i)U_iV_i)(
\sum_j K(\ww_j,\bb_j)U_j'V'_j)\right).\end{equation}

Now we claim that if $f_2-f_1=(1,0)$ then $$\sum_{i\in\gamma}
K(\ww_i,\bb_i)U_iV_i=z_0\frac{\partial P}{\partial z}(z_0,w_0)=z_0\alpha$$
and when $f_2-f_1=(0,1)$ then $$\sum_{i\in\gamma'}K(\ww_i,\bb_i)U_iV_i=w_0\frac{\partial P}{\partial w}(z_0,w_0)=w_0\beta,$$
and similarly for $f_4-f_3$. To see this, note first that the function
$K(\ww_i,\bb_j)U_iV_j$ is a function on edges which is a closed $1$-form, that is, a
divergence-free flow. In particular the sum $\sum_iK(\ww_i,\bb_i)U_iV_i$ is independent
of the choice of face path (in the same homology class).
We can therefore assume that the face paths $\gamma,\gamma'$ are equal to either
$\gamma_x$ or $\gamma_y$ according to whether they are horizontal or vertical.
Suppose for example $f_2-f_1=(1,0)$ and differentiate
$Q(z,w)K(z,w)=P(z,w)\cdot \mbox{Id}$ with respect to $z$ and evaluate at $(z_0,w_0)$: we get
$$Q_z(z_0,w_0)K(z_0,w_0)+Q(z_0,w_0)K_z(z_0,w_0)=P_z(z_0,w_0)\cdot \textup{Id}.$$
Applying $U$ from the right to both sides, using $K(z_0,w_0)U=0$ and
$Q(z_0,w_0)=UV^t$, and then multiplying both sides by $z_0$, this becomes
$$UV^tz_0K_z(z_0,w_0)U=z_0P_z(z_0,w_0)U.$$
However $z_0 K_z(z_0,w_0)=\sum_{\gamma_x} K(\ww,\bb)\ww\otimes \bb=
\sum_{i\in\gamma} K(\ww_i,\bb_i)\ww_i
\otimes \bb_i$, so we get
$$U\sum K(\ww_i,\bb_i)V_iU_i=z_0 P_z(z_0,w_0)U$$
and since $U\neq0$ the claim follows. The same argument works for $\gamma_y$.

Recall that $(x,y)$ was the translation between the fundamental domains containing
$g_p$ and $g'_q$.
In the sum (\ref{varsum}), let
$(x_1,y_1)\in\Z^2$ be the position of the fundamental domain of $g_p$  and $(x_2,y_2)$ that
of $g'_q$. Let $z_1=x_1\alpha z_0-y_1\beta w_0$ and
$z_2=x_2\alpha z_0-y_2\beta w_0$. The sum (\ref{varsum}) becomes (up to lower order terms)
$$-\frac1{2\pi^2}
\Re\int_{\phi(f_1)}^{\phi(f_2)}\int_{\phi(f_3)}^{\phi(f_4)}\frac{dz_1 dz_2}{(z_1-z_2)^2} $$
where $\phi$ is the linear map $(x,y)\mapsto x\beta w_0-y\alpha z_0.$
This integral evaluates to
$$\frac1{2\pi^2}
\Re\log\left(\frac{(\phi(f_4)-\phi(f_1))(\phi(f_3)-\phi(f_2))}{(\phi(f_4)-\phi(f_2))(\phi(f_3)-\phi(f_1))}
\right).$$
To approximate the height variance $\sigma(\tilde h(f_2)-\tilde h(f_1)),$
let $f_3$ be close to $f_1$ and $f_4$ close to $f_2$ (but still far enough apart on the scale
of the lattice so that the above approximations hold).
Then as $|f_2-f_1|\to\infty$ while $|f_3-f_1|$ and $|f_4-f_2|$ are remaining bounded,
the variance is
$$\frac1{\pi^2}\log|\phi(f_1)-\phi(f_2)|+o(\log|\phi(f_1)-\phi(f_2)|).$$
\end{proof}

\subsubsection{Case of a real node}\label{realnode}
In this section we show how to modify the above proof in case $P$ has a real node.
This happens when the two simple zeros $(z_0,w_0),(\bar z_0,\bar w_0)$ merge
into a single zero at one of the four points $(\pm1,\pm1)$, and
$P_z=P_w=0$ there. In this case the slope of the corresponding
EGM is integral but the amoeba does not have a complementary component,
 The component is reduced to a point (and is therefore not
``complementary'').

The canonical example of this behavior is the square grid with uniform
weights and a $2\times2$ fundamental domain; in this case
$P=4+z+z^{-1}+w+w^{-1}$, and there is a real node at $(z,w)=(-1,-1)$.

Since $K^{-1}(\bb,\ww)$ is a continuous function of the edge weights,
so is the variance of the height between two points.
When $P(z,w)$ has a node, at say $(z,w)=(1,1)$,
the polynomial $\tilde P(z,w)=P(e^{B_x}z, e^{B_y}w)$ has two simple zeros on $\T^2$
as long as $B_x,B_y$ are sufficiently close to but not equal to $0$.
So the height variance of $P$ is the limit of the height variances
of $\tilde P$ as $B_x,B_y\to0$.

In fact suppose without loss of generality that the node is at $(z,w)=(1,1)$ and $P$ has the expansion
$$P(z,w)=a(z-1)^2+b(z-1)(w-1)+c(w-1)^2+\dots,$$
where $a,b,c\in\R$ and $\dots$ denotes terms of order at least $3$. Then near the node
a point on $P$ satisfies either $z-1=\lambda (w-1)+O(w-1)^2$ or $z-1=\bar\lambda(w-1)+O(w-w_0)^2$ where
$\lambda,\bar\lambda$, which
are necessarily non-real, are the roots of $a+bx+c x^2=0$.

The proof of Theorem \ref{variance} is valid for $\tilde P$ except for one assertion,
where we ignored the cross terms in equation (\ref{cross}). Indeed,
when $z,w$ are each close to $1$ the cross terms do not oscillate.
However we will show that
$\sum_i K(\ww_i,\bb_i)U_i\bar V_i=o(|\alpha|+|\beta|)$ as $(z,w)$ tends to the node.
Along with the complex conjugate equation this proves that the
cross terms make no contribution.

The cross terms of \eqref{cross} give
\begin{equation}
-\frac1{2\pi^2}\frac{1}{|x\alpha-y\beta|^2}\left((\sum_{i}K(\ww_i,\bb_i)U_i\bar V_i)(\sum_jK(\ww_j,\bb_j)\bar U_j'V'_j)+(\text{conjugate})\right)\end{equation}
where recall that $\alpha=P_z,\beta=P_w$ are tending to $0$ at the node.
We first show that $\sum_i K(\ww_i,\bb_i)U_i\bar V_i=O(|\alpha|+|\beta|),$ and similarly for its
complex conjugate.
Recall the equation $QK=P\cdot \textup{Id}$. Differentiating with respect to $z$ we find
$$Q_zK+QK_z=P_z\cdot \textup{Id}.$$
At a point $(z_0,w_0)$ on $P$ we have $Q=UV^t$ and at $(\bar z_0,\bar w_0)$ we have
$Q=\bar U\bar V^t$. Applying $\bar U$ to the right and evaluating in the limit as $(z_0,w_0)$
tends to the node, we have
$$UV^tK_z\bar U=0$$ so that $0=V^tK_z\bar U=\sum_i K(\ww_i,\bb_i)U_i\bar V_i$ at the node.
Since $U,V$ can be chosen polynomial
in $z,w$ the quantity  $\sum_i K(\ww_i,\bb_i)U_i\bar V_i$ necessarily vanishes to order at least one
at the node (as do $\alpha$ and $\beta$).

Let $c_1$ be the limit at the node of $\frac{1}{\alpha}\sum_{i}K(\ww_i,\bb_i)U_i\bar V_i$ when $g_{p+1}-g_p=(1,0)$
and $c_2$ the same limit when $g_{p+1}-g_p=(0,1)$, so we may write
$$\lim\frac{1}{\alpha}\sum_{i}K(\ww_i,\bb_i)U_i\bar V_i=c_1dx+c_2dy$$ at the node.
The cross terms are then
$$-\frac1{2\pi^2}\frac{1}{|x-y\beta/\alpha|^2}\left((c_1dx_1+c_2dy_1)(\bar c_1dx_2+\bar c_2dy_2)+(\bar c_1dx_1+\bar c_2dy_1)(c_1dx_2+c_2dy_2)\right).$$
A short computation now shows that, since $\alpha/\beta\not\in\R$, this is not a closed $1$-form in $(x_1,y_1)$ or $(x_2,y_2)$
 unless $c_1=c_2=0$.

However since the height function differences $h(f_1)-h(f_2)$ do not depend on the path
$g_1,\dots,g_k$, the cross terms should necessarily be a closed $1$-form.
So $c_1=c_2=0$. We have proved

\begin{thm}\label{variancedegcpt}
Suppose $P$ has a real node $(z_0,w_0)=(\pm1,\pm1)$ on the unit torus,
and $$P(z,w)=a(z-z_0)^2+b(z-z_0)(w-w_0)+c(w-w_0)^2+\dots.$$
Then the height variance between two faces $f_1$ and $f_2$ is
$$\mbox{Var}[h(f_1)-h(f_2)]=\frac{1}{\pi}\log|\phi(f_1)-\phi(f_2)| +
o(\log|\phi(f_1)-\phi(f_2)| ),$$
where $\phi$ is the linear mapping $\phi(x+iy)= x z_0-y\lambda w_0$,
$\lambda$ being the root of $a+b\lambda+c\lambda^2=0$.
\end{thm}

\subsection{Gaseous phases (smooth non-frozen phases)}
When $(B_x,B_y)$ is in a bounded complementary component,
$P(e^{-B_x}z,e^{-B_y}w)$ has no zeros on the unit torus.
As a consequence $K^{-1}(\bb,\ww)$ decays exponentially fast in $|\bb-\ww|$.
\begin{prop}
The height variance $\sigma(h(f_1)-h(f_2))$ is bounded.
\end{prop}
\begin{proof}
The height difference $h(f_1)-h(f_2))$ can be measured along any path from
$f_1$ to $f_2$. Suppose a dual path from $f_1$ to $f_2$ is chosen so that
each edge crosses an edge of $G$ with black vertex on its left.
Then the height difference is a constant plus the sum of the indicator functions of the
edges on the path:  $h(f_1)-h(f_2)=\sum a_i-\bar a_i$.
Suppose we have two such paths $\gamma_1,\gamma_2$, consisting of edges $a_i$
and $b_j$ respectively,
which are close only near their endpoints.
The height variance is then the sum of the covariances of $a_i$ and $b_j$:
$$\sigma(h(f_1)-h(f_2))=\sum_{i,j}\Pr(a_i\text{ and }b_j)-\Pr(a_i)\Pr(b_j).$$
However these covariances are exponentially small except when both
$a_i$ and $b_j$ are near the endpoints $f_1$ or $f_2$.
In particular the above summation is a geometric series,
which has sum bounded independently of the distance between $f_1$ and $f_2$.
\end{proof}

\subsection{Loops surrounding the origin}

Smooth phases are characterized by their bounded height
variance or exponential decay of correlation. Here is another
characterization of smooth phases in terms of
loops in the union of two perfect matchings sampled
independently.

\begin{thm} A non-frozen EGM $\mu$ is smooth if and only if when two perfect matchings $M_1$ and
$M_2$ are chosen independently from $\mu$, there are almost surely only finitely many cycles in
$M_1 \cup M_2$ that surround the origin.  Similarly, a non-frozen EGM is rough if and only if when
two perfect matchings $M_1$ and $M_2$ are chosen independently from $\mu$, there are almost surely
infinitely many cycles in $M_1 \cup M_2$ that surround the origin. \end{thm}

\begin{proof} Since having infinitely many cycles surround the origin is a translation-invariant
event, it is clear that if $\mu$ is an EGM, then there are either $\mu$-almost-surely infinitely
many cycles or $\mu$-almost-surely finitely many cycles surrounding the origin.  If the 
former is
the case, it is easy to see that the variance is unbounded; to see this, simply use the fact that,
conditioned on the positions of the cycles, the two {\em orientations} of a given cycle (i.e.,
which alternative set of edges in the cycle belongs to which of the $M_i$) are equally likely, and
orientations of the cycles are independent of one another.  (Note that the orientation of the cycle
determines whether the height difference of the two height functions goes up or down when we cross
that cycle.)

Now suppose that  there are almost surely only finitely many cycles surrounding the origin.
 Lemma 8.4.3 of \cite{Sheffield} further implies that if two perfect matchings $M_1$ and
$M_2$ are sampled independently from $\mu(s,t)$, then the union $M_1 \cup M_2$ almost surely
contains no infinite paths. It follows that the height difference between the two height
functions is constant on the
infinite cluster of faces that are not enclosed in any loops.   Lemma 8.3.4 of \cite{Sheffield}
then implies that $\mu$ is smooth with respect to the (differently formulated but actually
equivalent) definition given in Chapter 8 of \cite{Sheffield}, and Lemmas 8.1.1 and 8.1.2 imply
that the height difference variances remain bounded in this case. \end{proof}

\section{Maximality of spectral curves}\label{smaxim}

\subsection{Harnack curves}

The characteristic polynomial $P(z,w)$ has real coefficients
and, hence, the spectral curve $P(z,w)=0$ is a real plane curve
(in fact, it is more natural to consider the spectral
curve as embedded in the toric surface corresponding
to the Newton polygon of $P$). While all smooth complex
curves of given genus are topologically the same,
the number and the configuration of the ovals of a real
plane curve can be very different. In particular, there
is a distinguished class of real plane curves, known
as \emph{Harnack} or \emph{maximal} curves, which have
the maximal number of ovals (for given Newton polygon)
in the, so to speak, best possible position. The
precise topological definition of a Harnack curve
can be found in \cite{Mikhalkin}; here we will use the
following alternative characterization of a Harnack
curve obtained in \cite{MR}. Namely, a curve $P(z,w)$
is Harnack if and only if the map from the curve
to its amoeba is $2$-to-$1$ over the amoeba interior
(except for a finite number of real nodes where it is $1$-to-$1$).
The main result of this section is the following

\begin{thm} \label{maximal} For any choice of
nonnegative edge weights the spectral curve
$P(z,w)=0$ is a Harnack curve.
\end{thm}

Harnack curves form a very special and much studied
class of curves. Several characterizations and many
beautiful properties of these curves can be
found in \cite{Mikhalkin,MR} (see also \cite{KO}). We will see
that several of them have a direct probabilistic
interpretation.

\subsection{Proof of maximality}

Maximality is an important property and several
proofs of it are available. In many respects,
it resembles the notion of total positivity \cite{GK,Kar}
and the proof given below exploits this analogy.
Proofs  of maximality based on different ideas can
be found in \cite{KO,KS}.

First observe that by the $2$-to-$1$ property, being
Harnack is a closed condition, hence it is enough
to prove that spectral curve is Harnack for a
generic choice of weights. Furthermore, it is
easy to see that we can obtain any periodic
planar bipartite graph as a limit case of the
periodically weighted hexagonal lattice when
some of the edge weights are zero. It is therefore,
enough to consider the case of generic periodic
weights on the hexagonal lattice with $n\times n$
fundamental domain.

\begin{figure}[hbtp]\psset{unit=0.5 cm}
  \begin{center}
    \begin{pspicture}(-2,0)(12,10)
    \rput(5,5){\scalebox{0.64}{\includegraphics{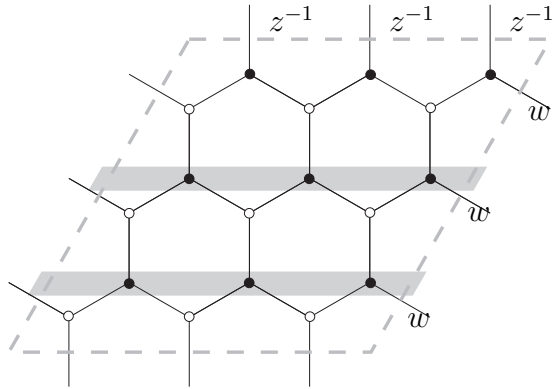}}}
    \rput[lb](4.7,9.4){$z^{-1}$}
    \rput[lb](7.9,9.4){$z^{-1}$}
    \rput[lb](11.1,9.4){$z^{-1}$}
    \rput[rt](9.0,1.9){$w$}
    \rput[rt](10.6,4.7){$w$}
    \rput[rt](12.2,7.4){$w$}
    \end{pspicture}
    \caption{\sl The solution to $K(z,w) f=0$ can be constructed
layer by layer}
    \label{ftransf}
  \end{center}
\end{figure}

By definition, $P(z,w)$ is a determinant of
an $n^2 \times n^2$ matrix $K(z,w)$ whose rows and columns
are indexed by the $n^2$ white and black
vertices in the fundamental domain.
We can also write $P(z,w)$ as a determinant of
an $n\times n$ matrix using transfer
matrices as follows. The equation $\det K(z,w)=0$
means that there exists a nonzero function $f$ on 
black vertices annihilated by the operator $K(z,w)$. 
We can construct such a function row by row 
as follows: given the values of $f$ on a
horizontal row of black vertices as in Figure \ref{ftransf},
the equation $Kf=0$ determines the values of
$f$ on the row below it. The corresponding linear
map is given by $-T(w)$, where $T(w)$ is the
transfer matrix of the form
\begin{equation}
T(w)=\begin{pmatrix}
a_1 & b_1\\
& a_2 & b_2 \\
&& a_3 & b_3 \\
&&&\ddots & \ddots\\
b_n \, w &&&& a_n
\end{pmatrix}\,, \quad a_i,b_i>0 \,.
\label{T(z)}
\end{equation}
Iterating this procedure
once around the period, we get a consistency 
relation, which gives 
$$
P(z,w)=\det\left(z-(-1)^n\, T_1(w)\cdots T_n(w)\right) \,.
$$

Suppose now that for some point $(x,y)\in \R^2$
the torus $\{|z|=e^x,|w|=e^y\}$ contains more
than 2 points of the curve $P(z,w)$. By changing
the magnetic field, we can assume that $y=0$.
That is, we can assume that for a pair of points
$(z_1,w_1)$ and $(z_2,w_2)$ on the spectral curve
we have
$$
|w_1|=|w_2|=1\,, \quad w_2 \ne w_1,\bar w_1\,, \quad |z_1|=|z_2| \,.
$$
Since being not Harnack is an open condition, we can
find a nearby curve for which both $w_1$ and $w_2$
are roots of unity. It is easy to see (a more
general statement is proved in \cite{KO}) that
we can achieve this by a small
perturbation of the dimer weights. So, we can assume
that $w_1^m=w_2^m=1$ for some integer $m$.

Taking $nm$ as the new horizontal period, we find that the
matrix
\begin{equation}
  \label{prodTi}
  M=T_1(1)\cdots T_n(1)
\end{equation}
has more than 2 eigenvalues of the same absolute value.
We will now show that this is impossible.

This follows from the following lemma which is
a version of a standard argument in the theory
of total positivity (cf.\ \cite{GK}).

\begin{lem}
  Suppose that all odd-size minors of a matrix $M$
are nonnegative and that there exists $k$ such that
all odd-size minors of $M^k$ are positive. Then
the eigenvalues of $M$ have the following form
$$
\lambda_1>|\lambda_2|\ge |\lambda_3| > |\lambda_4|\ge |\lambda_5| >
|\lambda_6|\ge |\lambda_7| > \dots  \,.
$$
\end{lem}

\noindent
Remark, in particular, that if $|\lambda_{2k}|> |\lambda_{2k+1}|$
then both of these eigenvalues are real.

\begin{proof}
Let us order the eigenvalues of $M$ so that
$$
|\lambda_1|\ge|\lambda_2|\ge |\lambda_3| \ge \dots \,.
$$
Since all matrix elements of $M$ are nonnegative and all matrix
elements of $M^k$ are positive, the Perron-Frobenius theorem
implies that $\lambda_1$ is simple, positive, and that
$$
\lambda_1>|\lambda_i|\,, \quad i>1\,.
$$
Now consider the action of the matrix $M$ in the third exterior
power $\Lambda^3{\mathbb R}^n$ of the original space ${\mathbb R}^n$.
The matrix elements of this action are the $3\times 3$ minors of
$M$ and, hence, Perron-Frobenius theorem again applies. The eigenvalues
of this action are the numbers
$$
\lambda_i \lambda_j \lambda_k \,, \quad 1\le i<j<k\le n \,.
$$
It follows that the number $\lambda_1 \lambda_2 \lambda_3$
is real, positive, and greater in absolute value than
$\lambda_1 \lambda_2 \lambda_i$ for any $i>3$. It follows that
$$
|\lambda_3| > |\lambda_4| \,.
$$
Iteration of this argument concludes the proof.
\end{proof}

It is immediate to see that any matrix of
the form $T_i(1)$ satisfies the hypothesis of the lemma.
Matrices with nonnegative minors of any given order
form a semigroup, so all odd-size minors of \eqref{prodTi}
are nonnegative. Genericity implies that all
odd-size minors of $M^k$ are positive for some large enough
$k$. This concludes the proof.

\subsection{Implications of maximality}

\subsubsection{Phase diagram of a dimer model}

The $2$-to-$1$ property immediately implies that:
\begin{itemize}
\item[(i)] the only singularities of the amoeba map are
folds over the boundary of the amoeba;
\item[(ii)] the boundary of the amoeba is the
image of the real locus of the spectral curve;
\item[(iii)] to any lattice point in the interior
of the Newton polygon corresponds either
a bounded component of the amoeba complement
or an isolated real node of the spectral curve.
In particular, the number of holes in the amoeba 
equals the geometric genus of the curve. 
\end{itemize}

For a general plane curve, more complicated
singularities of the amoeba map are possible,
which are then reflected in more complicated
singularities of the Ronkin function and, hence,
of its Legendre dual. In fact, we have used the
$2$-to-$1$ property in Section \ref{sphas} in the classification
of the phases of the dimer model. In this sense,
the probabilistic meaning of maximality is the
absence of any exotic phases with anomalous decay
of correlations.

Part (iii) implies that the gaseous phases persist
unless the corresponding component of the amoeba
complement shrinks to a point and a nodal singularity
develops. In particular, generically, the spectral
curve is smooth and all gaseous phases are present.
Conversely, if no gaseous phases are present then
the spectral curve has the maximal possible number
of nodes and hence is a curve of genus zero.
It is shown in \cite{KO} that the latter case
corresponds to isoradial dimers studied in \cite{Ke2}.
It is also shown in \cite{KO} that all Harnack
curves arise as spectral curves of some
dimer model.

To summarize, part (iii) implies the following:

\begin{thm}
The number of gaseous phases of a dimer model 
equals the genus of the spectral curve. 
For a generic choice of weights, the dimer
model has a gaseous phase for every lattice point in
the interior of the Newton polygon $N(P)$ and a frozen
phase for every lattice point on the boundary of $N(P)$.
\end{thm}

Part (ii) shows that the phase boundaries can
be easily determined. In fact, the amoeba
of a Harnack curve $P(z,w)=0$ can be defined by a
single inequality
\begin{equation}
  \label{ameq}
  \prod P(\pm e^x,\pm e^y) \le  0 \,,
\end{equation}
where the product is over all 4 choices of signs.
Indeed, it is clear that the product in \eqref{ameq}
changes sign whenever we cross the amoeba boundary
and, generically, is positive when $x$ or $y$ are
large. We remark that for a general, non-Harnack, curve it is
a rather nontrivial task to determine its
amoeba.

Using the interpretation of $P(\pm 1,\pm 1)$
as the expectation of $(\pm 1)^{h_x} (\pm 1)^{h_y} (-1)^{h_x h_y}$
with respect to the measure
$\mu_1$, see Section \ref{sfund}, we
arrive at the following:

\begin{thm}
The minimal free energy measure $\mu$ is smooth (that is, frozen or
gaseous) if and only if the $\mu_1$-measure of
one of the four $H_1(\T^2,Z/2)$ classes of matchings of $G_1$
exceeds $\frac12$, or can be made more than $\frac12$ by
an arbitrarily small perturbation of the dimer weights. 
\end{thm}

\subsubsection{Universality of height fluctuations}

In Theorem \ref{variance} we proved that
in a liquid phase the variance of the height function
difference grows like $\pi^{-1}$ times the
logarithm of the distance. The proof of that theorem
shows that the constant in front of the logarithm is
directly connected to the number of roots of the
characteristic polynomial on the unit torus. In
particular, maximality was used in the essential
way to show that this constant is always $\pi^{-1}$.
In fact, the universality of this constant is
equivalent to maximality.

\subsubsection{Monge-Amp\`ere equation for surface tension}

It follows from the results of \cite{MR} that the Ronkin
function $F$ of a Harnack curve satisfies the
following Monge-Amp\`ere equation
\begin{equation}
  \label{MA}
  \det
  \begin{pmatrix}
    F_{xx} & F_{xy} \\
    F_{yx} & F_{yy}
  \end{pmatrix}  = \frac1{\pi^2}\,,
\end{equation}
for any $(x,y)$ in the interior of the amoeba. By the
well-known duality for the  Monge-Amp\`ere equation,
this implies the analogous equation for the surface
tension function.

\begin{thm}\label{tMA} We have
\begin{equation}
  \label{MAs}
  \det
  \begin{pmatrix}
    \sigma_{xx} & \sigma_{xy} \\
    \sigma_{yx} & \sigma_{yy}
  \end{pmatrix}  = {\pi^2}\,.
\end{equation}
\end{thm}

It should be pointed out that in \cite{FPS} it
was argued that for certain class of random
surface models the equation \eqref{MAs} should
be satisfied at any \emph{cusp} of the surface
tension. It seems remarkable that in our case
\eqref{MAs} is satisfies not just at a cusp but
identically. For a general random
surface model, we only expect the left-hand side
of \eqref{MAs} to be positive, by strict
concavity.

The geometric meaning of the equations \eqref{MA}
and \eqref{MAs} is that the gradients of
$F$ and $\sigma$, which are mutually inverse
maps, are area-preserving, up to a factor. This
leads to another characterization of Harnack
curves as curves with amoebas of maximal
possible area for given Newton polygon $N(P)$,
namely $\pi^2$ times the area of $N(P)$. It
would be interesting to find a probabilistic
interpretation of this.

\subsubsection{Slopes and arguments}

Recall that by the results of Section \ref{ssurft}
the slope $(s,t)$
of the Ronkin function at a point $(B_x,B_y)\in \R^2$
equals the slope of the measure $\mu(B_x,B_y)$.
Suppose that the point $(B_x,B_y)$ is in the
interior of the amoeba and let $(z_0,w_0)$ be one
of its two preimages in the spectral curve.
Maximality connects the slope $(s,t)$ with the arguments
$z$ and $w$ as follows.

\begin{thm} \label{argslope} We have
$$
(s,t) = \pm\frac1{\pi}(\arg w_0,\arg z_0) \mod \Z^2 \,.
$$
\end{thm}

\begin{proof} We have
\begin{multline}\notag
  s=\frac{d}{dB_x}
\frac1{(2\pi i)^2}\iint_{\T^2} \log P(e^{B_x} z,e^{B_y} w)
\frac{dz}{z}\frac{dw}{w} \\
= \frac1{2\pi i}\int_{|w|=e^{B_y}}\left(\frac1{2\pi i}\int_{|z|=e^{B_x}}d\log P\right)\frac{dw}{w}\,.
\end{multline}
The inner integral counts,
for $w$ fixed, the number of zeros of $P(z,w)$ inside
$\{|z|=e^{B_x}\}$. As $w$ varies over the unit circle the number of zeros is locally constant
with unit jumps whenever a zero crosses the circle. These jump points are
precisely the points $w$ where $P(z,w)$ has a root on $\T^2$, that is precisely the two points
$w_0$ and $\bar w_0$ where $(z_0,w_0)$ and its conjugate are the unique zeros of $P$
on $\T^2$.
\end{proof}

\section{Random surfaces and crystal facets}

The goal of this section is to review some known facts about random surfaces in order to say
precisely what Theorem \ref{legendrethm} and Theorem \ref{mainmsr} imply about crystal facets and
random surfaces based on perfect matchings. We begin with some analytical results.

\subsection{Continuous surface tension minimizers} \label{continuousminimizers}

A standard problem of variational calculus is the following: given a bounded open domain $D \subset
\mathbb R^2$ and any strictly convex surface tension function $\sigma: \mathbb R^2 \rightarrow
\mathbb R$, find the function $f: D \rightarrow \mathbb R$ whose (distributional) gradient
minimizes the surface tension integral
$$
I(f) = \int_D \sigma (\nabla f(x))\,dx
$$
subject to the boundary condition that $f$ extends continuously to a function $f_0$ on the
boundary of $D$ and the volume condition that
$$
\int_{D} f(x) = B
$$ for some constant $B$.  In
our setting, we may also assume that $\sigma(u) = \infty$ whenever $u$ lies outside of the closure
of the Newton polygon $N(P)$, so that $f$ is necessarily Lipschitz.  The following result is well
known (see \cite{CKP}, \cite{Sheffield} for details and references).

\begin{prop} If there exists any $\tilde f$, satisfying prescribed volume and boundary constraints
and satisfying $I(\tilde f) < \infty$, then the surface tension minimizer $f$ is unique and its
gradient is almost everywhere defined. \end{prop}

Both $\nabla \sigma$ and $\nabla f$ are functions from $\mathbb R^2$ to $\mathbb R^2$.  The
Euler-Lagrange equation
for the functional $I(f)$ takes the following form:

\begin{prop} \label{eulerlagrange} Let $f$ be the surface tension minimizer described above.  Then
whenever $x \in D$, $f$ is $C^2$ at $x$, and $\sigma$ is smooth at $f(x)$, we have:
\begin{eqnarray} \label{eulerPDE} \textup{div}(\nabla \sigma \circ \nabla f(x)) &=& C \end{eqnarray}
for some constant $C$, which depends on $B$ and $f_0$.  If $f$ is also the minimal surface tension
function when the volume constraint is ignored, then (\ref{eulerPDE}) holds with $C=0$. \end{prop}

The archetypal solution to the equation in Lemma \ref{eulerlagrange} is the Legendre dual of
$\sigma$, which, in our setting, is the Ronkin function $F$. By construction
$$
\nabla \sigma \circ \nabla F = \textup{Id}
$$
throughout the amoeba---in
particular, the divergence is constant. 
By analogy with the case of the Ronkin
function we say that  $f$ has a \emph{facet} of slope $u$ if $\nabla f$ is
equal to $u$ on some open subset of $D$.

If $f_0$ is linear of any slope $u$ on the boundary of $D$, then it is easy to see that the minimal
surface tension function, ignoring the volume constraint, is linear of slope $u$, has (trivially) a
facet of slope $u$, and satisfies (\ref{eulerPDE}) with $C=0$.  If we require, however, that $C \not
= 0$, so that the
volume constraint exerts some non-zero amount of pressure (upward or downward) on
the surface, then Lemma \ref{eulerlagrange} and Theorem \ref{aboutsigma} imply that the slope of
any facet of $f$ must be a lattice point inside $N(P)$. In other words, the facet slopes
of the Ronkin function $F$ represent all possible facet slopes of the dimer model.

\subsection{Concentration inequalities for discrete random surfaces}
In this section, we aim to show that the surface tension minimizing shapes and facets described in
Section \ref{continuousminimizers} are approximated by perfect-matching-based discrete random
surfaces.  First, suppose that $D$ is a domain in $\mathbb R^2$, that $f_0$ is continuous Lipschitz
function defined on $\partial D$, and that $f$ is a surface tension minimizer (with no volume
constraint) which agrees with $f_0$ on the boundary.

Next, denote by $\frac{1}{n} G$ the infinite graph $G$ whose embedding into $\R^2$ has been
re-scaled by a factor of $1/n$.  For example, when $G = \Z^2$, then $\frac{1}{n} G$ is a grid
mesh that is $n$ times finer than $G$.  Suppose that $D_n$ is a sequence of simply connected
subgraphs of $\frac{1}{n}G$ that {\it approximate $(D,f)$ from the inside} in the sense that
\begin{enumerate}
\item The embedding of $D_n$ is contained in $D$ for all $n$.
\item The Hausdorff distance between the boundary of $D_n$ and the boundary of $D$ tends to zero in $n$.
\item Each $D_n$ admits at least one perfect matching, and for these matchings, the height functions
$h_n$ on the boundary of $D_n$ are such that $\sup |h_n - f|$ tends to zero
in $n$ (where $h_n$ is treated as a function on a subset of the points in $D$, by letting $h_n(x)$
denote the height at the face of $D_n$ containing $x$). \end{enumerate}

For each $n$, define $\nu_n$ to be the Boltzmann measure on perfect matchings $M_n$ of $D_n$.
Intuitively, we would expect that for sufficiently large values of $n$, the normalized function
$h_n/n$ will closely approximate the continuous function $f$ with high $\nu_n$ probability.  A
version of this statement is proved in \cite{CKP} in the case $G = \Z^2$.  Analogous but more
general statements (in the form of \emph{large deviations principles}) are discussed in Chapter 7 of
\cite{Sheffield} (see the paragraphs on Lipschitz potentials in Sections 7.3 through 7.5). In
both \cite{CKP} and \cite{Sheffield}, the results imply that, for a fixed value of $\epsilon$,
$\nu_n \{ |h_n/n - f| > \epsilon \}$ tends to zero exponentially in $n^2$.  Also, if $x \in D$ is a
point at which $\nabla f(x) = u$, we would expect the local statistics of $h_n$, near the point
$x$, to be those of the Gibbs measure $\mu_{u_1, u_2}$.  More precise versions of this statement
can be found in Chapter 7 \cite{Sheffield}.  The issue of weighting by enclosed volume is addressed
briefly in Section 7.5 of \cite{Sheffield}.

\end{document}